\documentclass[twocolumn,twocolappendix]{aastex631}

\usepackage{comment}
\usepackage{bm}
\usepackage{multirow}
\usepackage{braket}
\usepackage{appendix}
\usepackage{CJK}

\usepackage{amsmath}

\shorttitle{Hot Jupiter Occurrence Rate}
\shortauthors{Miyazaki and Masuda}

\begin{document}

\title{
Evidence that the Occurrence Rate of Hot Jupiters around Sun-like Stars Decreases with Stellar Age
}

\author[0000-0001-9818-1513]{Shota Miyazaki}
\affiliation{Institute of Space and Astronautical Science, Japan Aerospace Exploration Agency, 3-1-1 Yoshinodai, Chuo, Sagamihara, Kanagawa 252-5210, Japan}
\author[0000-0003-1298-9699]{Kento Masuda}
\affiliation{Department of Earth and Space Science, Graduate School of Science, Osaka University, 1-1 Machikaneyama, Toyonaka, Osaka 560-0043, Japan}

\begin{abstract}
We investigate how the occurrence rate of giant planets (minimum mass $> 0.3\, M_\mathrm{Jup}$) around Sun-like stars depends on the age, mass, and metallicity of their host stars.
We develop a hierarchical Bayesian framework to infer the number of planets per star (NPPS) as a function of both planetary and stellar parameters. The framework fully takes into account the uncertainties in the latter by utilizing the posterior samples for the stellar parameters obtained by fitting stellar isochrone models to the spectroscopic parameters, Gaia DR3 parallaxes, and 2MASS $K_{\rm s}$-band magnitudes adopting a certain bookkeeping prior.
We apply the framework to 46 Doppler giants found around a sample of 382 Sun-like stars from the California Legacy Survey catalog that publishes spectroscopic parameters and search completeness for all the surveyed stars.
We find evidence that the NPPS of hot Jupiters (orbital period $P=1$--$10\,\mathrm{days}$) decreases roughly in the latter half of the main sequence over the timescale of $\mathcal{O}(\mathrm{Gyr})$, while that of cold Jupiters ($P=1$--$10\,\mathrm{yr}$) does not.
Assuming that this decrease is real and caused by tidal orbital decay, the modified stellar tidal quality factor $Q^\prime_\star$ is implied to be $\mathcal{O}(10^6)$ for a Sun-like main-sequence star orbited by a Jupiter-mass planet with $P\approx 3\,\mathrm{days}$.
\end{abstract}

\section{Introduction \label{sec:intro}}
\defcitealias{Rosenthal+2021}{R21}
Quantifying the occurrence rate of hot Jupiters (HJs) is important for understanding their formation and evolution \citep{Dawson+2018}, and this has been the focus of many previous works.  
The rates estimated for Sun-like stars are all roughly around 1\%, but the results based on different survey samples show differences of marginal significance \citep[e.g.,][]{Wright+2012, Howard+2012}. 
This may partly be due to the unaccounted correlation between the HJ occurrence and certain properties of the stars, which may vary among the stellar samples that were used to derive the occurrence rates. 
Stellar metallicity shows a strong correlation with the occurrence rate of giant planets and has long been recognized as one such important variable \citep{Santos+2004, Fischer+2005, Guo+2017}.
More recently, it has been argued that stellar environments may also play an important role. For example, \citet{Brucalassi+2016, Brucalassi+2017} estimated the HJ occurrence rate in the solar-metallicity, solar-age open cluster M67 to be $5.7^{+5.5}_{-3.0}\%$, which may be higher than that around field stars. \citet{Winter+2020} also claimed that the occurrence rate of close-in planets including HJs is enhanced around stars that are clustered in the position-velocity phase space, although the key driver of this correlation is still debated \citep[e.g.,][]{Kruijssen+2021}.
\citet{Mustill+2022}, in particular, pointed out that the correlation found by \citet{Winter+2020} can instead be a manifestation of the age dependence: stars in phase-space overdensity are also kinematically cold and tend to be younger, and so the correlation may arise if HJs prefer younger hosts. 

This paper primarily focuses on the age dependence of the HJ occurrence, which has been less well explored than the dependence on the other parameters. 
Given that the orbital periods of HJs are often shorter than the rotation periods of their host stars, the possibility that HJs may spiral into the central star due to tides raised on it has been discussed since the discovery of 51 Peg b \citep{Rasio+1996, Valsecchi+2014b, Valsecchi+2014a}. 
While it has been difficult to theoretically predict the orbital decay timescale due to the complex nature of tidal dissipation \citep{Ogilvie+2014}, there is growing observational evidence in favor of ongoing decays. 
Long-term transit monitoring of WASP-12b \citep{Maciejewski+2016, Patra+2017, Yee+2020} and Kepler-1658b \citep{Vissapragada+2022} indicate that the orbital periods as measured directly by successive transits are decreasing. 
The rapid rotation reported for HJ hosts may also be a signature of stellar spin-up associated with orbital decay, i.e., angular momentum transfer from the orbit to spin \citep[e.g.,][]{Penev+2018, 2021ApJ...919..138T}. 
Even the detection of an infrared transient associated with the engulfment of a short-period planet has been reported \citep{De+2023}.
Of particular relevance to our paper is the work by \citet{Hamer+2019}, who showed that Sun-like stars known to host HJs have a smaller Galactic velocity dispersion than their control stars without known planets, and those with known cold Jovian planets. 
They interpreted this difference as evidence that HJ hosts are on average younger than their control stars, suggesting that a non-negligible fraction of HJs are tidally disrupted by their host stars during the main sequence lifetime.

However, it still remains quantitatively unclear how the orbital decay drives the evolution of the entire population of HJs. 
The two systems with detected decay, for example, may well be atypical cases showing the quickest decays that are easiest to detect, perhaps aided by enhanced tidal dissipation due to host stars evolving off the main sequence \citep[e.g.,][]{SW2013, Weinberg+2017}. 
The study of \citet{Hamer+2019} does not relate the velocity dispersion with the absolute age scale. 
Even after such a conversion, it will not be straightforward to interpret the inferred mean age of the HJ hosts, because it depends not only on the occurrence rate of HJs as a function of age but also on the age distribution of the stars from which these planets have been drawn (see also Section~\ref{ssec:npps}). 
It is difficult to test whether the latter is the same as their control stars or stars with cold Jupiters because of the inhomogeneous nature of their HJ sample.
The HJ occurrence around red giants reported by \citet{Temmink+2023} may not even support the qualitative conclusion of \citet{Hamer+2019}, although the estimate based on four HJs still has a large uncertainty. 

Here we attempt to infer the HJ occurrence rate as a function of the stellar age, which directly probes what fraction of HJs is affected by orbital decay on what timescale, and potentially provides insights into the physics of tidal dissipation and formation of HJs \citep[e.g.,][]{MK2016, OL2018}. 
Unlike in \citet{Hamer+2019}, this requires the knowledge of not only planet-hosting stars but also of all the stars from which these planets have been detected, as well as the age estimates for all of them.
To meet these requirements, here we focus on the California Legacy Survey (CLS) sample \citep[][hereafter \citetalias{Rosenthal+2021}]{Rosenthal+2021} --- which published the information of all the surveyed stars including the search completeness --- and leverage isochronal ages that can be estimated homogeneously for all the stars. 
This is still a challenging task because main-sequence stars change little with age and thus the uncertainty of the isochronal age is typically large due to degeneracy with other stellar properties and also varies depending on the evolutionary phase of individual stars \citep{Soderblom2010}. 
In this paper, we develop a hierarchical Bayesian framework that allows us to properly handle the large (but statistically well-defined) uncertainty of the isochrone-based age, as well as the degeneracy with stellar mass and metallicity. 

The structure of this paper is as follows. 
In Section \ref{sec:sample}, we describe our isochrone analysis of the CLS stars and construct a subsample of the CLS stars that we use for the occurrence rate analysis.
In Section \ref{sec:framework}, we present our Bayesian framework for inferring the occurrence rate of planets as a function of both planetary properties (mass, orbital period, radius, etc.) and stellar properties (mass, metallicity, age, etc.). 
In Section \ref{sec:inference}, we apply the framework to the Sun-like star sample defined in Section~\ref{sec:sample} and infer the occurrence rate of hot and cold Jupiters as a function of stellar parameters.
In Section \ref{sec:discussion}, we discuss constraints on the stellar tidal quality factor and implications for other survey results. 
Section~\ref{sec:summary} summarizes the paper.

\begin{figure}[t!]
    \centering
    \includegraphics[bb= 0 0 398 406,scale=0.6]{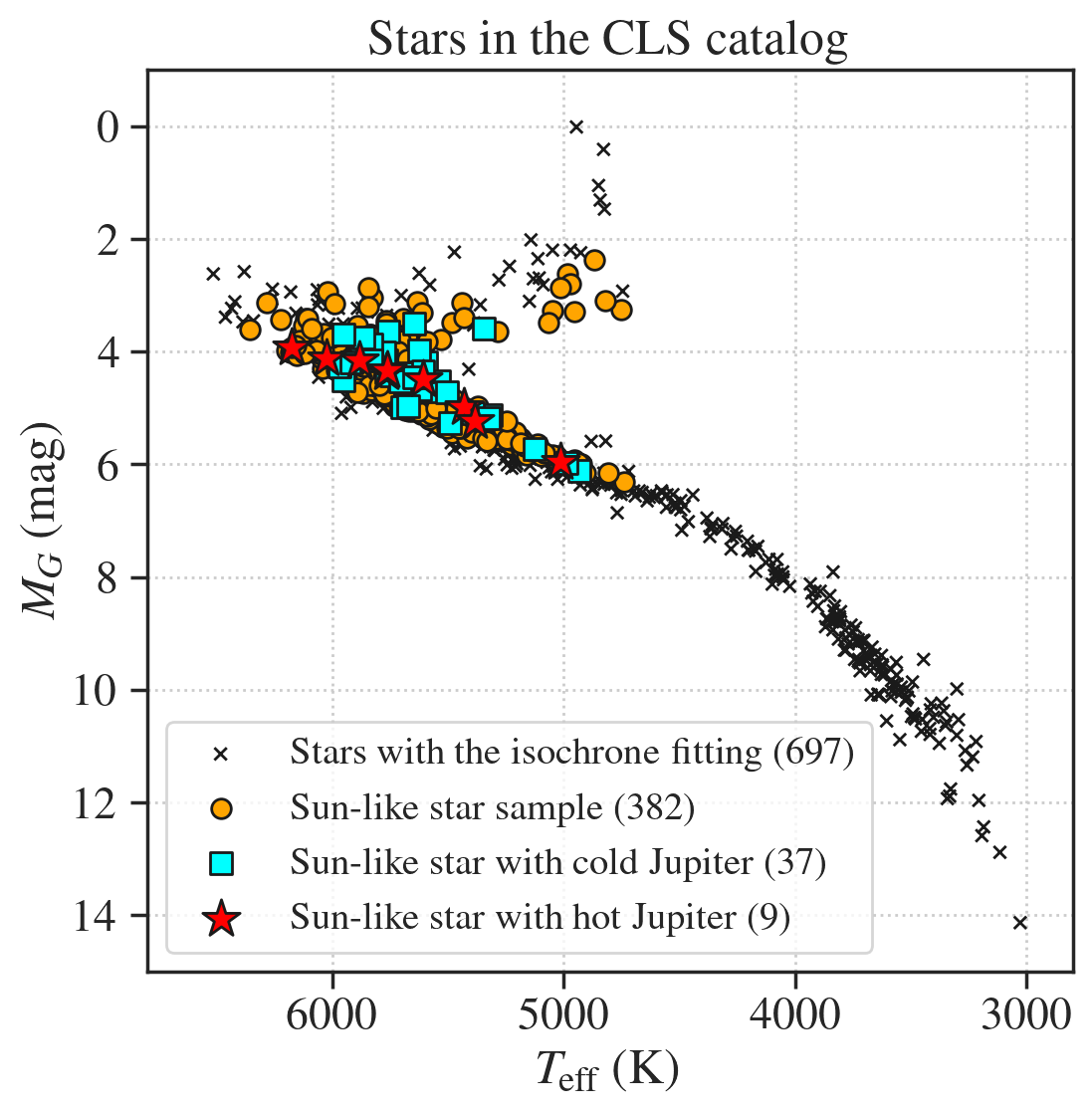}
    \caption{
    Black crosses show the absolute Gaia $G$-band magnitudes $M_{G}$ and the effective temperatures $T_{\rm eff}$ of 697 CLS stars for which we derived isochrone parameters.
    382 stars in our Sun-like sample defined in Section \ref{sec:sample} are shown as orange circles.
    Among them, 37 stars (cyan squares) host cold Jupiters, and 9 stars (red stars) host hot Jupiters.
    }
    \label{fig:Teff_MG}
\end{figure}

\begin{figure*}[t!]
    \centering
    \includegraphics[bb=0 0 566 403,scale=0.9]{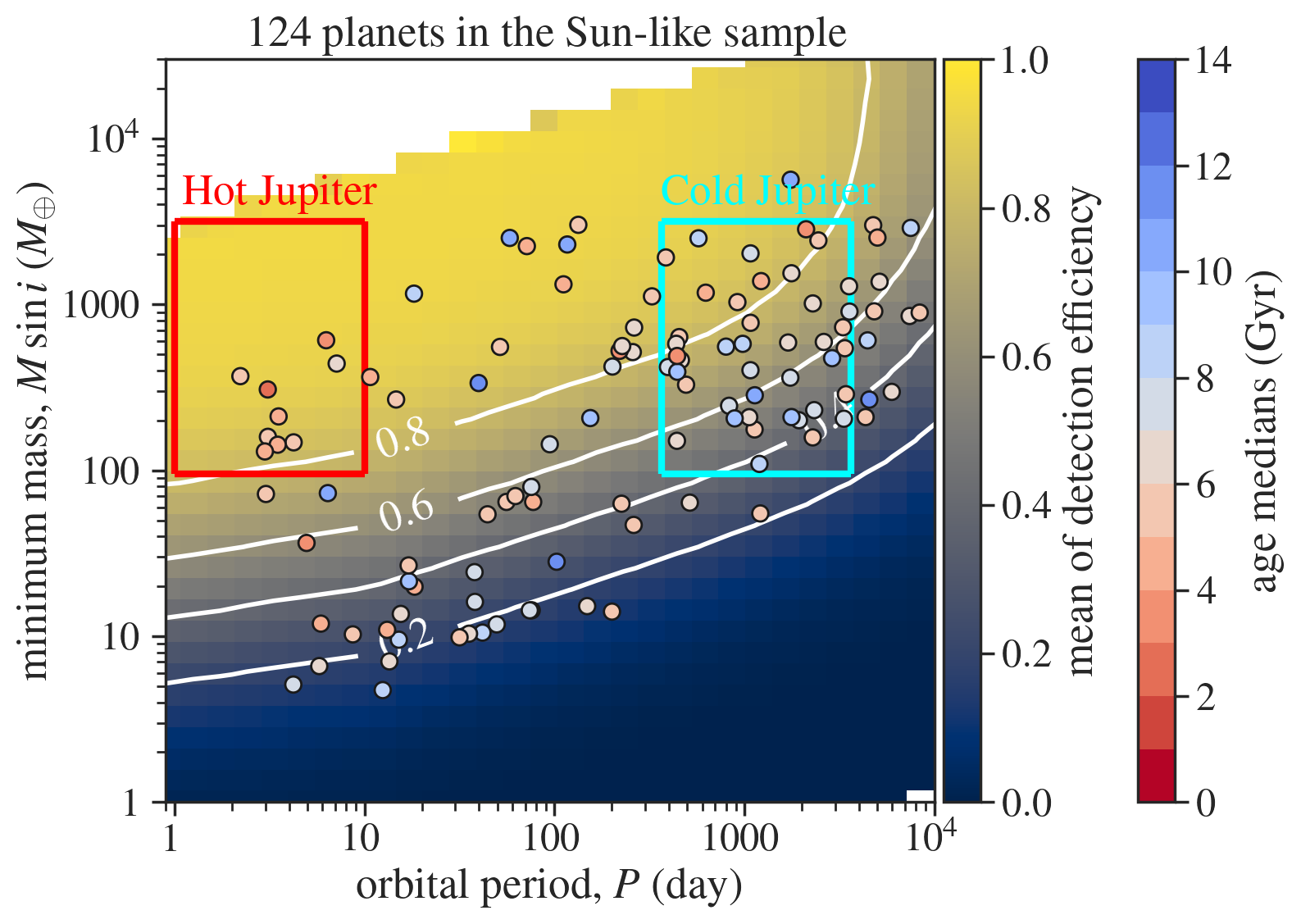}
    \caption{
    The minimum RV masses $M\sin{i}$ and orbital periods $P$ of planets in our Sun-like star sample (filled circles).
    Planets are colored by median values of ages in the posterior PDFs that are obtained by the stellar isochrone fitting described in Section \ref{sec:isochrone}.
    The red and cyan rectangles show the definition of hot and cold Jupiters in our analysis, respectively.
    The background color map and white contours represent the mean detection efficiencies (survey completeness) of planets in the Sun-like star sample calculated from the injection-recovery simulation results published in \citet{Rosenthal+2021}.
    }
    \label{fig:planets_DE}
\end{figure*}

\begin{figure*}[t!]
    \centering
    \includegraphics[bb=0 0 732 430,scale=0.7]{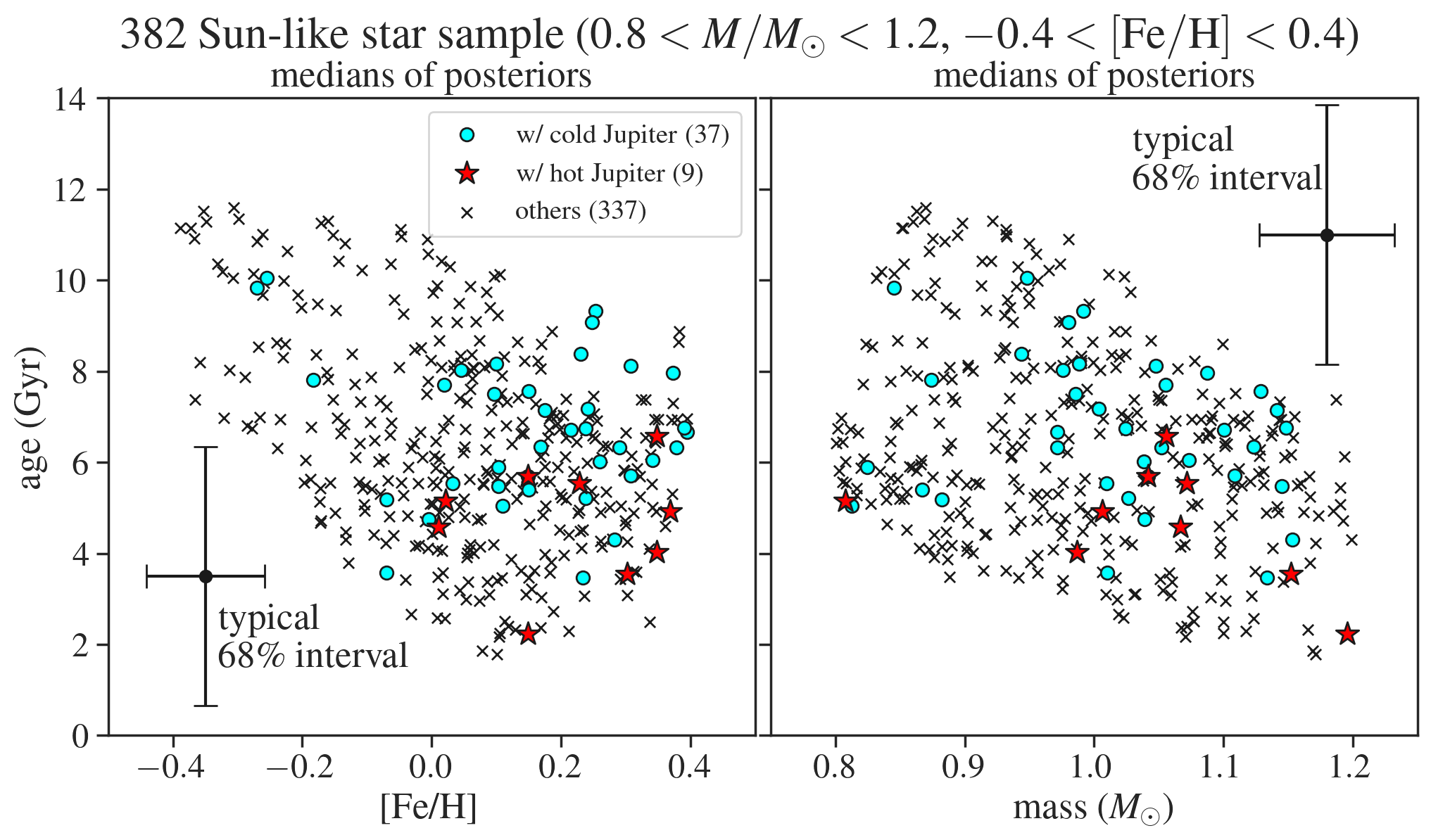}
    \caption{
    The result of the isochrone fitting for the 382 stars in the Sun-like sample.
    {\bf (Left panel)}: Medians of the posterior distributions in the plane of metallicity [Fe/H] and age.
    The red stars and cyan circles represent hosts of hot Jupiter (HJ) and cold Jupiter (CJ), respectively.
    The typical 68\% credible interval in each posterior distribution is shown at the bottom left.
    {\bf (Right panel)}: Same as the left panel, but for the stellar mass and age.
    }
    \label{fig:Final_sample}
\end{figure*}

\begin{figure}[t!]
    \centering
    \includegraphics[bb=0 0 520 520,scale=0.48]{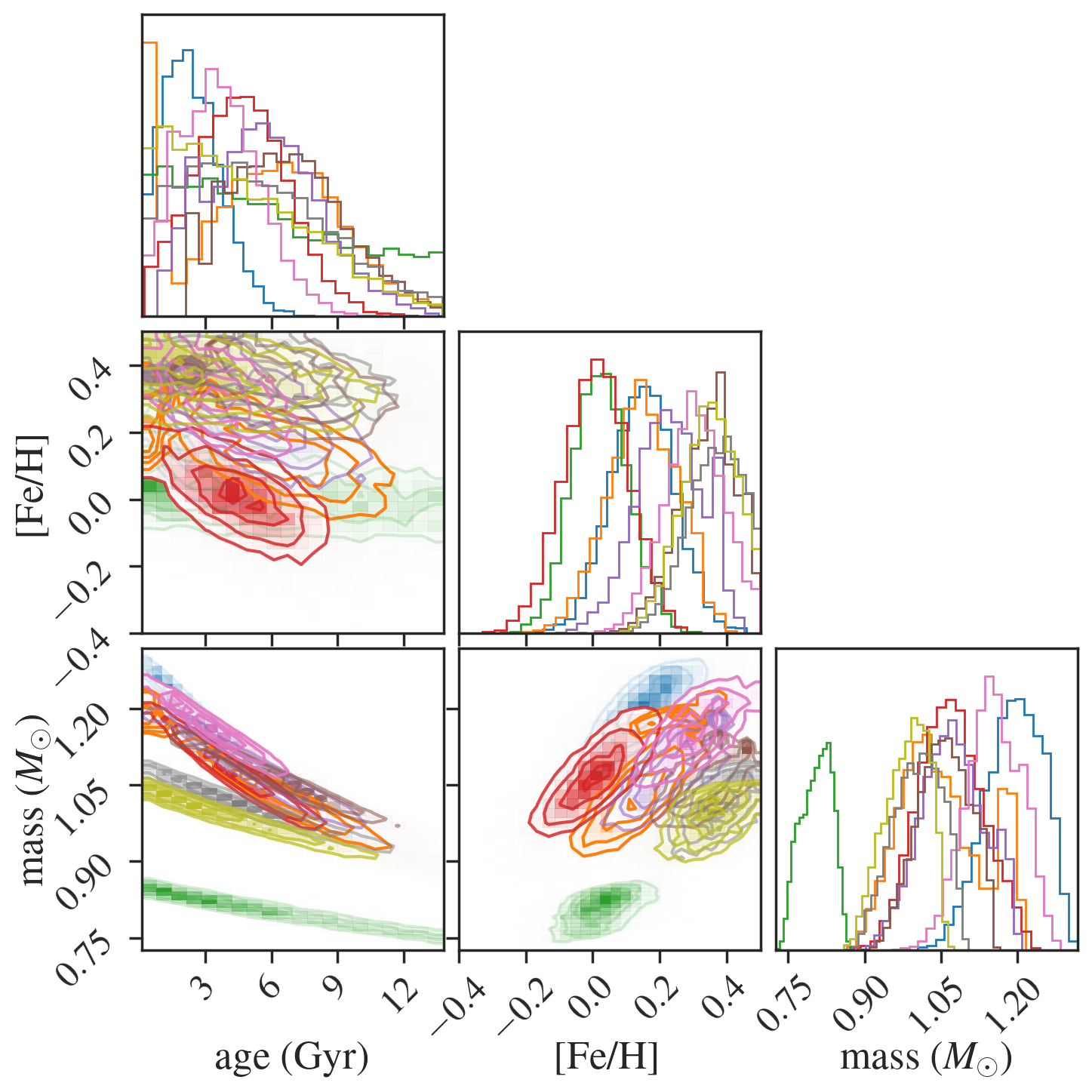}
    \caption{
    Corner plots for the posterior samples of the parameters of 9 HJ hosts in the Sun-like sample.
    The inferred age, mass, and metallicity are highly correlated.
    This is also the case for the other stars in the Sun-like sample.
    This figure was created using {\tt corner.py} \citep{corner}.
    }
    \label{fig:MCMC_HJ}
\end{figure}

\begin{figure}[t!]
    \centering
    \includegraphics[scale=0.6,bb=0 0 408 330]{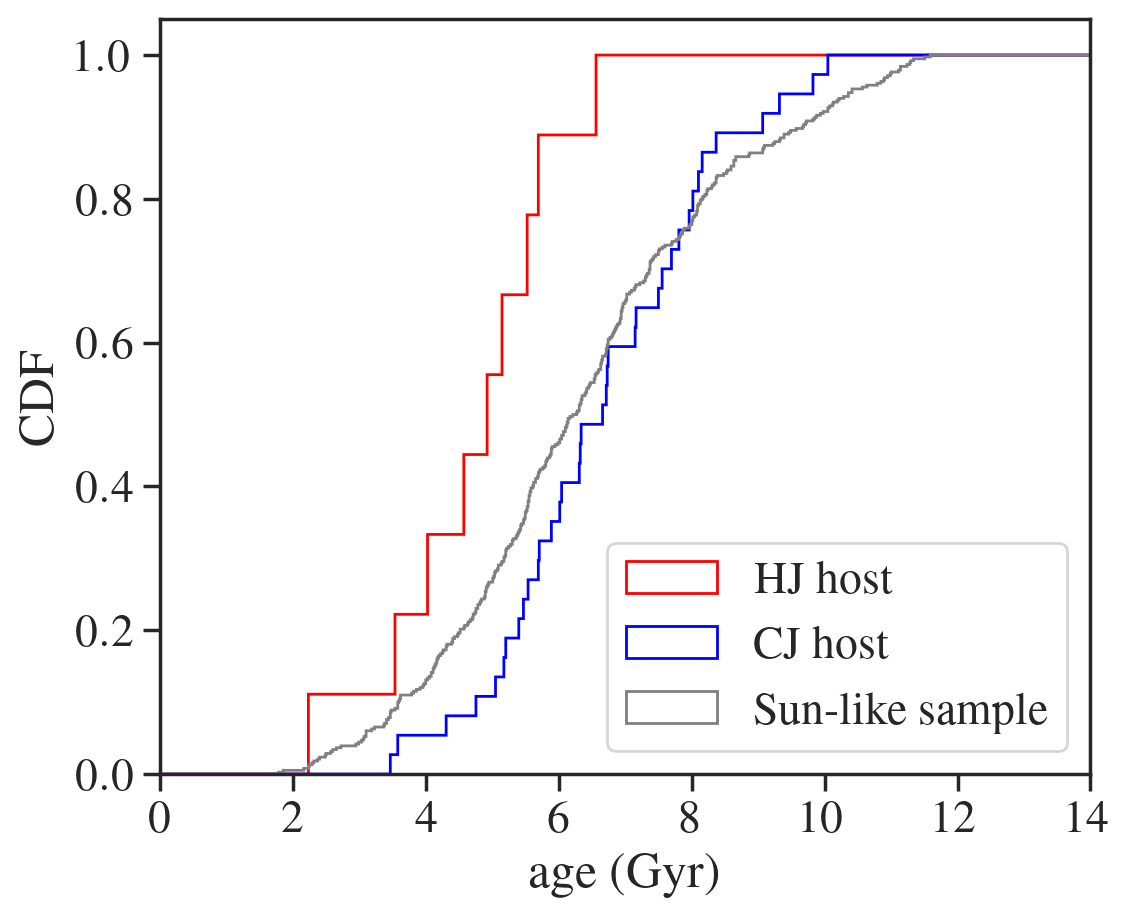}
    \caption{
    Cumulative distribution functions (CDFs) of the ages for the HJ hosts (red), CJ hosts (blue), and all the stars (gray) in the Sun-like sample, respectively. Here the ``age" for each star is evaluated as the median of the marginalized posterior distribution.
    }
    \label{fig:cdf_age}
\end{figure}

\section{Sample Description \label{sec:sample}}
\subsection{The California Legacy Survey}

The CLS catalog presented by \citetalias{Rosenthal+2021} contains 179 planets detected around 719 nearby FGKM dwarfs based on the radial velocity (RV) data from the California Planet Search team \citep{Howard+2010}. 
\citetalias{Rosenthal+2021} made various data related to the CLS survey available, including spectroscopic parameters of the stars, properties of detected planets derived from RVs, and planet detection efficiencies (completeness),\footnote{https://github.com/leerosenthalj/CLSI} which makes the catalog an ideal resource for occurrence rate studies.
In the following subsections, we will define a subset of Sun-like stars from the CLS catalog for our occurrence analysis. 
The stars in our sample are shown in Figure \ref{fig:Teff_MG} with filled orange circles.
We will use the spectroscopic parameters provided in the catalog for stellar isochrone fitting in Section \ref{sec:isochrone}, and utilize the planetary parameters and detection efficiencies for inferring the planet occurrence rate in Section \ref{sec:inference}.

\subsection{Isochrone fitting \label{sec:isochrone}}

We fit the observed effective temperature $T_{\rm eff}$, iron metallicity [Fe/H], $K_{\rm s}$-band magnitude $K_{\rm s}$, and parallax $\varpi$ of the CLS stars with the MIST models \citep{Paxton+2011, Paxton+2013, Paxton+2015, Choi+2016}, and derive physical parameters of the stars including mass and age. 
The outcome is the samples drawn from the joint posterior distribution for the stellar parameters of each star, obtained using the {\tt jaxstar} software\footnote{https://github.com/kemasuda/jaxstar} \citep{Masuda2022, jaxstar}. 
The samples will be used for inferring the occurrence rate of planets in Section \ref{sec:framework}.

We use the measurements of $T_{\rm eff}$ and [Fe/H] from the CLS catalog, which were obtained in \citetalias{Rosenthal+2021} by applying the {\tt Specmatch} \citep{Petigura2015} algorithm to the Keck-HIRES high-resolution spectra and using the {\tt Isoclassify} \citep{Morton2015, Huber+2017, Berger+2020} software package.\footnote{Strictly speaking, it is ideal here to use the direct outputs of {\tt Specmatch} that do not rely on stellar models.
We implicitly assume that this minor inconsistency does not affect the inferred age dependence given the errors we assigned for these parameters. 
}
We assume Gaussian errors of 100~K and 0.1~dex for $T_{\rm eff}$ and [Fe/H], respectively. 
We exclude five stars without $T_{\rm eff}$ measurements and eight stars with [Fe/H]$<-1$ in the CLS catalog.
We collect the parallax measurements ($\varpi$) from Gaia DR3 \citep{Gaia+2016, Gaia+2022} by cross-referencing using their coordinates and magnitudes. 
We correct the zero-points of the parallax values following the recipe of \citet{Lindegren+2021} and their error bars following \citet{El-Badry+2021}. 
We obtain $K_{\rm s}$-band magnitudes from the Two Micron All Sky Survey \citep[2MASS;][]{Skrutskie+2006} using the Gaia DR3 source IDs and correct the measurements for extinction using the dust map of {\tt Bayestar17} \citep{Green+2019}, although the corrections turned out to be negligible in most cases due to their proximity and near-infrared magnitudes.
We removed stars without the 2MASS IDs.
Six stars in the CLS catalog do not have $K_{\rm s}$-band measurements, so we instead use $H$-band 2MASS magnitudes for these stars. In the end, all the necessary information was found for 697 stars.

We then perform isochrone fitting for the 697 stars. By fitting, we mean drawing samples from the joint posterior probability density function (PDF) of stellar parameters $\bm{\theta}$, $p_0(\bm{\theta}|D)$, conditioned on the data $D$ and a certain prior PDF $p_0(\bm{\theta})$ for ``bookkeeping."\footnote{The inference of the occurrence rate will not depend explicitly on the prior adopted here; see Section~\ref{sec:framework}.}
Given the measured values of $\bm{y}^{\rm obs}=\{T_{\rm eff}, \mathrm{[Fe/H]}, \varpi, K_{\rm s}\}$ and their errors $\bm{\sigma}^{\rm obs}$, the likelihood function $\mathcal{L}(D|\bm{\theta})$ based on the assumption that errors follow Gaussian distributions and are independent is:
\begin{eqnarray}
    \mathcal{L}(D|\bm{\theta}) = \prod_i \frac{1}{\sqrt{2\pi}\sigma^{\rm obs}_i}\exp{\left[-\frac{1}{2} \left(\frac{y^{\rm obs}_i-y_i(\bm{\theta})}{\sigma^{\rm obs}_i}\right)^2\right]},
\end{eqnarray}
where $y_i(\bm{\theta})$ is computed (in a deterministic manner) by linearly interpolating the MIST grids of $(t_\star, \mathrm{[Fe/H]}, e)$ and by $\varpi=1/d$,
where $t_\star$, $e$, and $d$ represent the stellar age, equivalent evolutionary phase \citep[EEP;][]{Dotter+2016}, and distance to the star, respectively. 
We then sample from:
\begin{eqnarray}
    p_0(\bm{\theta}|D) \propto \mathcal{L}(D|\bm{\theta})\,p_0(\bm{\theta}).
\end{eqnarray}
We assume that the prior PDF $p_0(\bm{\theta})$ is separable as $p_0(\bm{\theta}) = p_0(t_\star,\mathrm{[Fe/H]},e)\,p_0(d)$ and set:
\begin{eqnarray}
    p_0(t_\star,\mathrm{[Fe/H]},e) \propto \left|\frac{\partial (t_\star,\mathrm{[Fe/H]},M_\star)}{\partial (t_\star,\mathrm{[Fe/H]}, e)}\right|
\end{eqnarray}
in order that the prior PDF is constant in the parameter space of $(t_\star, \mathrm{[Fe/H]}, M_\star)$ where valid stellar isochrone models exist \citep{Morton2015, Masuda2022}.
The priors for age, [Fe/H], and EEP are bounded within $(0.1, 13.8)$ Gyr, $(-0.5, 0.5)$, and $(0, 600)$, respectively. 
For the distance prior $p_0(d)$, we adopt an exponentially decreasing volume density prior with a length scale of 1.35 kpc \citep{Bailer-Jones+2015, Astraatmadja+2016}.
We obtain 10,000 posterior samples for each of the 697 stars in the CLS catalog (black crosses in Figure~\ref{fig:Final_sample}).
For all the parameters, the Gelman--Rubin convergence diagnostics $\hat{R}$ \citep{Gelman1992} computed by splitting a single chain were $\hat{R}<1.01$ in most stars and $\hat{R}<1.2$ in all the stars.

\subsection{The Sun-like Star Sample}

In this work, we focus on planets around Sun-like stars, for which models are more accurately calibrated than for lower- and higher-mass stars \citep{Soderblom+2010}.
We first removed 25 poorly-fitted stars whose marginalized posterior PDFs do not contain any of the measured values within 90\% credible intervals, resulting in 672 stars.
We construct the subsample of Sun-like stars by choosing the stars whose mass and [Fe/H], as evaluated using the median of the posterior samples, satisfy $0.8<M^{\rm med}_{\star}/M_{\odot}<1.2$ and $-0.4<\mathrm{[Fe/H]}^{\rm med}<0.4$, respectively, and then we are left with 382 stars.
Typical $68\%$ credible intervals for each posterior are approximately $0.05\,M_\odot$ in mass, $0.09\,{\rm dex}$ in [Fe/H], and $2.8\,{\rm Gyr}$ in age. 
A previous comparison with asteroseismic stars demonstrated no significant bias larger than these uncertainties for the stars in the above mass-metallicity range \citep{Masuda2022}.

The Sun-like sample contains 124 planets shown in Figure~\ref{fig:planets_DE}. We focus on the following two classes of planets:
\begin{itemize}
\item Hot Jupiter (HJ): Planets with orbital period $P$ ranging from 1--10$\;{\rm days}$ and minimum mass $M\sin{i}$ between 0.3--10$\;M_{\rm Jup}$.
\item Cold Jupiter (CJ): Planets with orbital period $P$ between 1--10$\;{\rm years}$ (corresponding to approximately 1--5 au in semi-major axis) and minimum mass $M\sin{i}$ in the range of 0.3--10$\;M_{\rm Jup}$.
\end{itemize}
The corresponding regions of the parameter space are shown by red and cyan rectangles in Figure~\ref{fig:planets_DE}. 
Also shown by the background color map and white contours are the mean detection efficiencies as computed using the results of injection-recovery simulation reported in \citetalias{Rosenthal+2021}.
The minimum masses of the planets $M\sin{i}$ and the detection efficiencies are recalculated using the host masses $M_\star$ derived from isochrone fitting.
For most of the HJs and CJs as defined here, the uncertainties in the minimum mass and orbital period are smaller than the bin size adopted for displaying the detection efficiency.

\subsection{Initial Investigation of the Age Dependence: A Simple Statistical Test and its Problems}\label{ssec:simple_analysis}

In Figure \ref{fig:planets_DE}, each planet is color-coded based on the age of the host star, here evaluated as the median of the marginal posterior. 
Albeit with large uncertainties (see also Figures~\ref{fig:MCMC_HJ}), this figure already hints that stars hosting HJs (all having reddish colors) tend to be younger than those with CJs (mixture of blue and red).\footnote{One may think that the lower limit of the HJ's mass, $M\sin{i}=0.3M_{\rm Jup}$, can impact our result because a relatively older planet-host exists at $(P, M\sin{i})=(6\;{\rm day},70M_{\oplus})$ in Figure \ref{fig:planets_DE}. We confirmed our conclusions remained unchanged when the planets down to $M\sin{i}=0.1\, M_{\rm Jup}$ were considered as HJs.}
The same trend is also seen in Figure \ref{fig:Final_sample}, which displays the distribution of the posterior medians of $(\mathrm{[Fe/H]},M_\star,t_\star)$ in the Sun-like sample: the stars hosting HJs (red stars) are in the lower parts of the figures at a given [Fe/H] (left panel) or mass (right panel), compared to stars with CJs (cyan circles) or stars without known planets (black crosses).

We perform a simple statistical test for the age distribution as an attempt to quantify this difference further and to highlight its issues as well.
Figure~\ref{fig:cdf_age} shows the cumulative distribution functions of ages (median of marginal posterior) for all stars in the sample (gray), stars hosting HJs (red), and stars with CJs (blue). 
We tested the null hypothesis that two of the subsamples are drawn from the same population using the Anderson--Darling test for $k$ samples and found the $p$-values of 0.022, 0.18, and $1.6\times10^{-3}$ when comparing all-stars vs. HJ hosts, all-stars vs. CJ hosts, and HJ hosts vs. CJ hosts, respectively. These results confirm the above visual impression that HJ hosts are younger than CJ hosts as well as all surveyed stars.

We note, however, that interpreting this kind of statistical test is by no means straightforward. 
First, this does not take into account heterogeneous uncertainties in the age estimates. In general, the uncertainty varies from star to star for a variety of reasons; this may be due to the difference in the precision of the spectroscopic parameters that depends on the signal-to-noise of the observed spectrum, or due to the difference in other stellar parameters (e.g., mass) that affect the age precision. When this is the case, the above test may give low $p$-values even for two populations of stars with the same true age distributions; the distribution of the medians of the posteriors conditioned on a certain prior (i.e., the estimated ages) could still be different.\footnote{Consider an extreme case where there exist two populations of stars with the same age distributions and the ages can be inferred extremely precisely for one population, and only very poorly for the other. Then the distribution of the {\it estimated} ages will be very similar to the true distribution in the former but will be totally different for the latter, depending on how the estimate is given, and so the distributions of the estimated ages will look very different.} 
Second, the apparent difference in the age distributions could be caused by the difference in other stellar parameters, such as the mass, that are correlated with age. If the occurrence rate of HJs increases with the stellar mass more rapidly than that of CJs, for example, then the HJ hosts will be younger than CJs {\it even if their occurrences have the same age dependence at a given mass}, because more massive stars are on average younger. 
One conceptually straightforward solution is to control the other stellar parameters in the sample by, for example, focusing on subsets of stars with similar masses and metallicities and computing the occurrence rates separately. When this is not practical --- as in our case --- one needs to consider the distribution of all the relevant stellar parameters in the sample and infer the dependence of the occurrence rate on these parameters simultaneously.
Such an analysis must take into account the fact that the constraints on stellar parameters are often given in a degenerate manner; for example, the isochronal mass and age are strongly degenerate (Figure~\ref{fig:MCMC_HJ}). The degeneracy in the inferred stellar parameters should not be confused with the correlation that actually exists in the stellar sample.

The rest of this paper describes our attempts to analyze the age difference taking into account the above issues associated with large and heterogeneous uncertainties in the degenerate stellar parameters.
The following Section \ref{sec:framework} presents a framework for doing so.

\section{A General Framework for Inferring the Number of Planets per Star \label{sec:framework}}
\subsection{Definition of Planet Occurrence Rate}\label{ssec:npps}

Two types of ``occurrence rates" have been mainly discussed in the literature: 
(1) the number of planets per star \citep[NPPS; e.g.][]{Youdin+2011, Fulton+2021}, and (2) the fraction of stars with planets \citep[FSWP; e.g.][]{Fischer+2005, Johnson+2010}. 
In this work, we discuss the NPPS as a function of both planet and stellar properties.

Let $\bm{z}$ and $\bm{x}$ represent the physical parameters of stars (mass, metallicity, age, etc.) and planets (mass, orbital period, radius, etc.), respectively. 
We define the NPPS function $f(\bm{x}|\bm{z})$ so that, around a star with given $\bm{z}$, the expected number of planets that fall within a small volume of the parameter space $d\bm{x}$ around $\bm{x}$ is  \citep[e.g.,][]{Tabachnik+2002, Youdin+2011}:
\begin{eqnarray}\label{eq:def_f}
    d\overline{n}_{\rm p} = 
    f(\bm{x}|\bm{z}) d\bm{x}.
\end{eqnarray}
Thus $f(\bm{x}|\bm{z})$ denotes the expected number of planets per star and per $\mathrm{d}\bm{x}$, but {\it not} per $d\bm{z}$. 

By integrating over a certain region $\mathcal{P}$ in the $\bm{x}$-space, which defines the ``planet" of consideration, we obtain the expected number of planets with $\bm{x}\in\mathcal{P}$ per star, in a manner that depends on the stellar properties:
\begin{eqnarray}\label{eq:np_z}
        \overline{n}_{\bm{x} \in \mathcal{P}}(\bm{z})
        = \int_\mathcal{P} f(\bm{x}|\bm{z}) d\bm{x}.
\end{eqnarray}
Further integrating over $\bm{z}$, the expected NPPS around $N_\star$ stars whose properties $\bm{z}$ follow the probability distribution $p_\star(\bm{z})$ is:
\begin{align}
\notag
    \overline{n}_{\bm{x}\in\mathcal{P}, \bm{z}\sim p_\star(\bm{z})}
    &= {1 \over N_\star} \int_\mathcal{S} \overline{n}_{\bm{x} \in \mathcal{P}}(\bm{z}) N_\star p_\star(\bm{z})\,d\bm{z} \\
    \label{eq:np}
    &= \int_\mathcal{P}\int_\mathcal{S} f(\bm{x}|\bm{z})\,p_\star(\bm{z})\,d\bm{z} \,d\bm{x},
\end{align}
where $\mathcal{S}$ denotes the domain of $\bm{z}$.

We emphasize again that $f(\bm{x}|\bm{z})$ does not have the unit of $1/\bm{z}$, and neither does $\overline{n}_{\bm{x} \in \mathcal{P}}(\bm{z})$. The PDF $p_\star(\bm{z})$ does, and satisfies $\int_\mathcal{S} p_\star(\bm{z})\,d\bm{z}=1$. 
One cannot integrate out $\bm{z}$ in $f(\bm{x}|\bm{z})$ without specifying $p_\star(\bm{z})$. 
This means the properties of all the surveyed stars need to be specified to relate $f$ with the observed number of planets. 
To put it the other way, the parameter distribution of the surveyed stars necessarily needs to be specified/inferred for inferring the NPPS as a function of stellar parameters.

\subsection{Inferring the NPPS from the Data}

Consider a planet survey in which the total number of stars is $N_\star$.
We assume that the survey provides the following data sets:
\begin{itemize}
    \item $\bm{D}\equiv\{D_j\}^{N_\star}_{j=1}$: This data has information about the stellar properties $(\bm{z}_j)$ for each star.
    In our case, this is the stellar data used for isochrone fitting.
    \item $\bm{H}=\{H_j\}^{N_\star}_{j=1}$: This data set includes the parameters of planets around each star. 
    For the $j$-th star with detected planets, the set $H_j$ contains the parameters for all the detected planets, represented as $H_j=\{\bm{x}^{(1)}_j,\bm{x}^{(2)}_j,.. .,\bm{x}^{(n_j)}_j\}$, where $n_j$ is the number of detected planets orbiting the $j$-th star.  
    If no planets are detected, the set will be empty. In our application, $\bm{x}$ consists of the minimum mass and orbital period of the planet.
\end{itemize}
We parameterize the stellar distribution $p_\star(\bm{z})$ and the NPPS function $f(\bm{x}|\bm{z})$ by the sets of parameters $\bm{\alpha}$ and $\bm{\gamma}$, respectively. 
We then aim to determine the joint posterior PDF of these parameters, given the data $\bm{D}$ and $\bm{H}$:
\begin{eqnarray}\label{eq:sample}
    p(\bm{\alpha},\bm{\gamma}|\bm{D},\bm{H}) &\propto& \mathcal{L}(\bm{\alpha},\bm{\gamma})\,p(\bm{\alpha},\bm{\gamma}),
\end{eqnarray}
where $\mathcal{L}(\bm{\alpha},\bm{\gamma})$ is the likelihood function and $p(\bm{\alpha},\bm{\gamma})$ is the prior PDF for $\bm{\alpha}$ and $\bm{\gamma}$.
We assume that the data for each star are independent when conditioned on $\bm{\alpha}$ and $\bm{\gamma}$. 
Then the likelihood is represented as the product of the probabilities of the data for each star:
\begin{eqnarray}
    \mathcal{L}(\bm{\alpha},\bm{\gamma}) &=& \prod^{N_\star}_{j=1}p(D_j,H_j|\bm{\alpha},\bm{\gamma}).
\end{eqnarray}
For the $j$-th star, the likelihood may be evaluated as:
\begin{eqnarray}\label{eq:pDHag}
    p(D_j,H_j|\bm{\alpha},\bm{\gamma}) &=& 
    \int p(D_j,H_j|\bm{\alpha},\bm{\gamma},\bm{z}_j)\,p(\bm{z}_j|\bm{\alpha}, \bm{\gamma})d\bm{z}_j\nonumber\\
    &=& \int p(D_j|\bm{z}_j)\,p(H_j|\bm{\gamma},\bm{z}_j)p_\star(\bm{z}_j|\bm{\alpha})d\bm{z}_j.\nonumber\\
\end{eqnarray}
In the second line, we assume that $D_j$ and $H_j$ are independent when conditioned on $\bm{z}_j$, and that $D_j$ does not depend on $\bm{\alpha}$ nor $\bm{\gamma}$ when conditioned on $\bm{z}_j$.
We compute $p(H_j|\bm{\gamma},\bm{z}_j)$ assuming that the number of detected planets in a given small partition $\Delta_l$ of the parameter space follows the Poisson distribution, and that the detections in different partitions are made independently. 
From Equation~(\ref{eq:def_f}), the expected number of planets in a given partition is $\eta_j(\bm{x}_l)f(\bm{x}_l|\bm{z}_j)\Delta_l$, where $\eta_j(\bm{x})$ is the detection efficiency of planets as a function of $\bm{x}$ for the $j$-th star. 
Note that $\eta_j(\bm{x})$ is not a density function per $\bm{x}$. 
Then $p(H_j|\bm{\gamma},\bm{z}_j)$, the probability to find one planet in partitions around $\bm{x}_j^{(1)}, \dots, \bm{x}_j^{(n_j)}$, but none elsewhere, is \citep{Tabachnik+2002, Youdin+2011}:
\begin{eqnarray}
    &&p(H_j|\bm{\gamma},\bm{z}_j) \nonumber\\
    &=& \underbrace{\left(\prod_{l,\bm{x}_l\in H_j}\left[\eta_j(\bm{x}_l)f(\bm{x}_l|\bm{\gamma},\bm{z}_j)\Delta_l\right]\exp{\left[-\eta_j(\bm{x}_l)f(\bm{x}_l|\bm{\gamma},\bm{z}_j)\Delta_l\right]} \right)}_{\rm for\;partitions\;with\;detected\;planets} \nonumber\\
    &&\times \underbrace{\left(\prod_{l,\bm{x}_l\notin H_j}\exp{\left[-\eta_j(\bm{x}_l)f(\bm{x}_l|\bm{\gamma},\bm{z}_j)\Delta_l\right]} \right)}_{\rm for\;partitions\;without\;detected\;planets}\nonumber\\
    &\propto& \left(\prod_{\bm{x}_j\in H_j} \eta_j(\bm{x}_j)f(\bm{x}_j|\bm{\gamma},\bm{z}_j)\right) \exp{\left[-\int\eta_j(\bm{x})f(\bm{x}|\bm{\gamma},\bm{z}_j)d\bm{x} \right]}.\nonumber \\
\end{eqnarray}
Substituting this expression into Equation (\ref{eq:pDHag}), we find
\begin{eqnarray}\label{eq:DjHj}
    p(D_j,H_j|\bm{\alpha},\bm{\gamma}) &=& \int \;p(D_j|\bm{z}_j)\,p_\star(\bm{z}_j|\bm{\alpha}) \left(\prod_{\bm{x}_j\in H_j} \widetilde{f}(\bm{x}_j|\bm{\gamma},\bm{z}_j)\right) \nonumber\\
    &&\times \exp{\left[-\int \widetilde{f}_j(\bm{x}|\bm{\gamma},\bm{z}_j)d\bm{x} \right]} d\bm{z}_j,
\end{eqnarray}
where we defined $\widetilde{f}_j(\bm{x}|\bm{z}) = \eta_j(\bm{x})f(\bm{x}|\bm{z})$.

In this work, we model the distribution of stellar parameters in the sample, $p_\star(\bm{z})$, as follows:
\begin{eqnarray}
    p_{\star}(\bm{z}|\bm{\alpha})=\sum^{K}_{k=1} \exp{(\alpha_k)}\;\bm{1}_{k}(\bm{z})\label{eq:star_dist}.
\end{eqnarray}
Here, $k$ represents the index for bins that divide the domain of $\bm{z}$, and $\bm{1}_{k}(\bm{z})$ is a step function that returns 1 if the $k$-th bin contains $\bm{z}$, and 0 otherwise.
The term $\exp{(\alpha_k)}$ represents the height of the $k$-th bin, and must satisfy $\sum_k \exp{(\alpha_k)}\Delta_k=1$ with $\Delta_k$ being the volume of the $k$-th bin, so that $p_\star$ is a normalized PDF.
In this form, we assume that $p_{\star}(\bm{z}|\bm{\alpha})$ is constant within a given bin, and $\alpha_k$ represents the log probability density in the $k$-th bin.
Using this expression for $p_\star(\bm{z}|\bm{\alpha})$, Equation (\ref{eq:DjHj}) can be rewritten as:
\begin{eqnarray}
    p(D_j,H_j|\bm{\alpha},\bm{\gamma}) &=& \sum^{K}_{k=1} \exp{(\alpha_k)} L_{j,k}(\bm{\gamma}), 
\end{eqnarray}
where
\begin{eqnarray}
    L_{j,k}(\bm{\gamma}) &=& \int d\bm{z}_j \, p(D_j|\bm{z}_j) \bm{1}_k(\bm{z}_j) \left(\prod_{\bm{x}_j\in H_j} \widetilde{f}_j(\bm{x}_j|\bm{\gamma},\bm{z}_j)\right)\nonumber\\
    &\times& \exp{\left[-\int \widetilde{f}_j(\bm{x}|\bm{\gamma},\bm{z}_j)d\bm{x} \right]} \nonumber. 
\end{eqnarray}

Following \citet{Hogg+2010}, we evaluate the integral of $L_{j,k}$ via an importance sampling using the samples from the posterior PDF $p_0(\bm{z}|D_j)$ conditioned on a certain bookkeeping prior $p_0(\bm{z})$, as we obtained in Section~\ref{sec:isochrone}. Namely,
considering Bayes' theorem as 
\begin{equation}
    p(D|\bm{z}) = \frac{p_0(\bm{z}|D)p_0(D)}{p_0(\bm{z})},
\end{equation}
we approximate $L_{j,k}$ as:
\begin{eqnarray}\label{eq:L_approx}
    L_{j,k}(\bm{\gamma}) &\approx& \frac{p_0(D_j)}{M} \sum^{M}_{m=1} \frac{\bm{1}_k(\bm{z}^{(m)}_j)}{p_0(\bm{z}^{(m)}_j)} \left(\prod_{\bm{x}_j\in H_j} \widetilde{f}_j(\bm{x}_j|\bm{\gamma},\bm{z}_j)\right) \nonumber\\
    &\times& \exp{\left[-\int \widetilde{f}_j(\bm{x}|\bm{\gamma},\bm{z}_j)d\bm{x} \right]},\;\;\;\bm{z}^{(m)}_j\sim p_0(\bm{z}|D_j). \nonumber\\
\end{eqnarray}
Here, $M$ represents the number of samples drawn from the posterior PDF, and we can drop the constant factor $p_0(D)$ that is irrelevant to the inference.
We note that the inference does not depend explicitly on the choice of the bookkeeping prior $p_0(\bm{z})$.  
This is because the evaluation of $L$ only involves $p_0(\bm{z}|D)/p_0(\bm{z}) \propto p(D|\bm{z})$, which is the likelihood function that depends on the data alone, but not on $p_0(\bm{z})$ used to draw posterior samples in the isochrone fitting of individual stars.

Finally, the joint posterior PDF in Equation~(\ref{eq:sample}) is evaluated as
\begin{eqnarray}\label{eq:summary}
    p(\bm{\alpha},\bm{\gamma}|\bm{D},\bm{H}) = \prod^{N_\star}_j p(D_j,H_j|\bm{\alpha},\bm{\gamma})\,p(\bm{\alpha},\bm{\gamma}).
\end{eqnarray}
In this work, we perform posterior sampling using Hamiltonian Monte Carlo and No-U-Turn Sampler \citep{Duane+1987, Betancourt2017} as implemented in {\tt NumPyro} \citep{phan2019composable,bingham2019pyro}. 
The code is implemented using {\tt JAX} \citep{Bradbury+2021}.

\section{Planet Occurrence Rates around the CLS Sun-like Stars\label{sec:inference}} 
\begin{figure*}[t!]
    \centering
    \includegraphics[bb=0 0 809 404,scale=0.63]{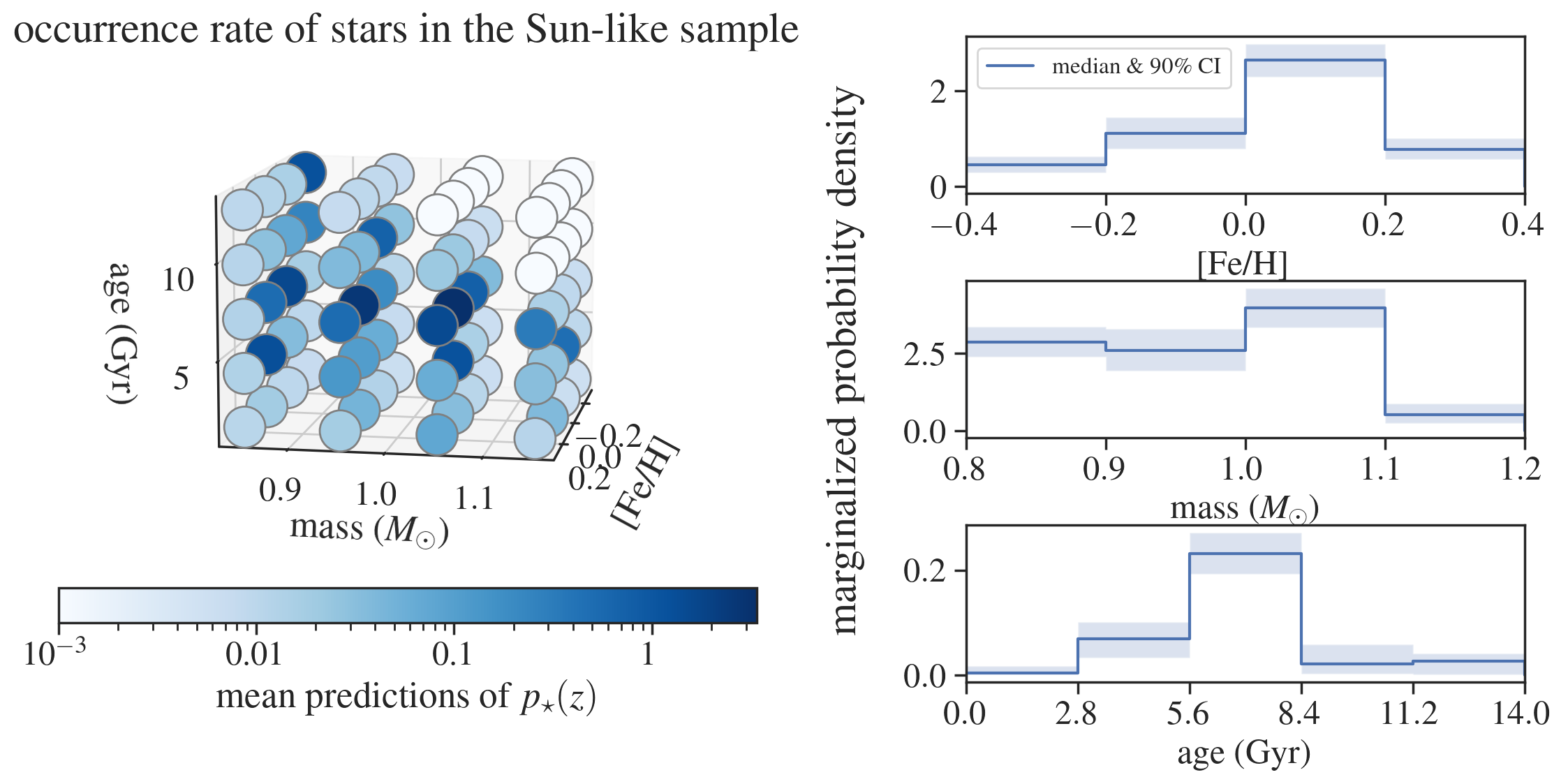}
    \caption{
    The stellar parameter distribution $p_\star(\bm{z})$ inferred for the Sun-like sample using the histogram model in Section~\ref{subsec:histogram}.
    ({\bf Left}): The 3D view of the posterior in the parameter space of $(\mathrm{[Fe/H]}, M_\star, t_\star)$, where the circles represent the mean values of $p_\star(\bm{z})$ at the center of each grid.
    ({\bf Right}): The panels show the marginalized posterior PDFs for [Fe/H], mass, and age (top to bottom). 
    The solid lines and translucent regions represent the medians and 90\% equal-tail credible intervals (CIs) in each bin.
    }
    \label{fig:occ_nonpara}
\end{figure*}

\begin{figure*}[t!]
    \centering
    \includegraphics[bb=0 0 1011 431,scale=0.5]{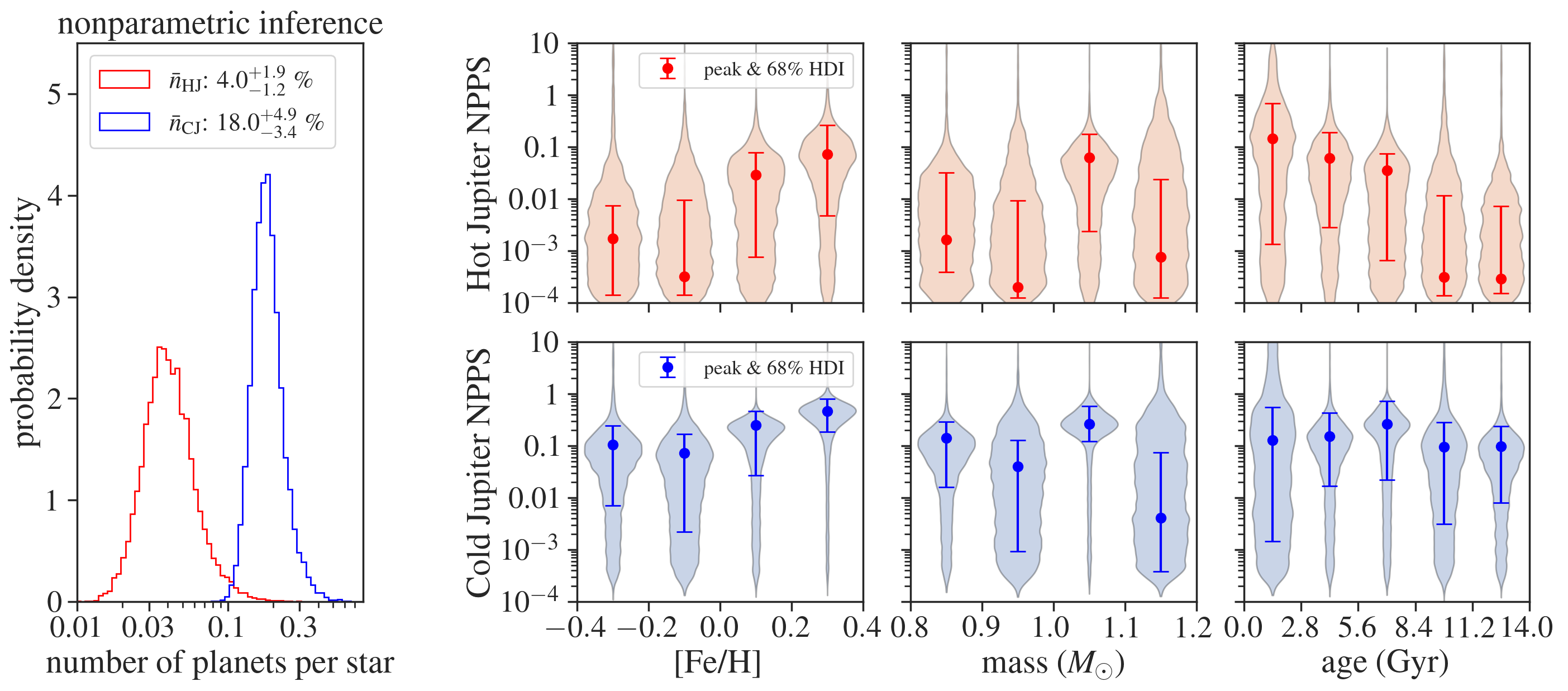}
    \caption{
    The NPPS function $\bar{n}_p(\bm{z})$ inferred for the Sun-like sample using the histogram model in Section~\ref{subsec:histogram}.
    ({\bf Left}): The posterior PDFs of the NPPS for HJ (orange) and for CJ (blue). These are the numbers of planets per star in the searched sample, computed taking into account the stellar parameter dependence of the occurrence rate as well as the distribution of the stellar parameters in the sample; see Equation (\ref{eq:np}). 
    ({\bf Right}): The posterior PDFs for the NPPS of HJs (top) and CJs (bottom) as functions of [Fe/H], mass, and age are shown as violin plots, whose widths represent the probability densities.
    The error bars represent their peaks (modes) and 68\% highest density intervals (HDIs).
    This is essentially the NPPS as a function of $\bm{z}$ as given in Equation~\ref{eq:np_z} but has been marginalized for the stellar parameters that are not shown in the horizontal axis as in Equation~(\ref{eq:np}) (see also the main text and Equation~(\ref{eq:violin})).
    }
    
    \label{fig:NPPS_nonpara}
\end{figure*}

\begin{figure}[t!]
    \centering
    \includegraphics[bb=0 0 634 397,scale=0.38]{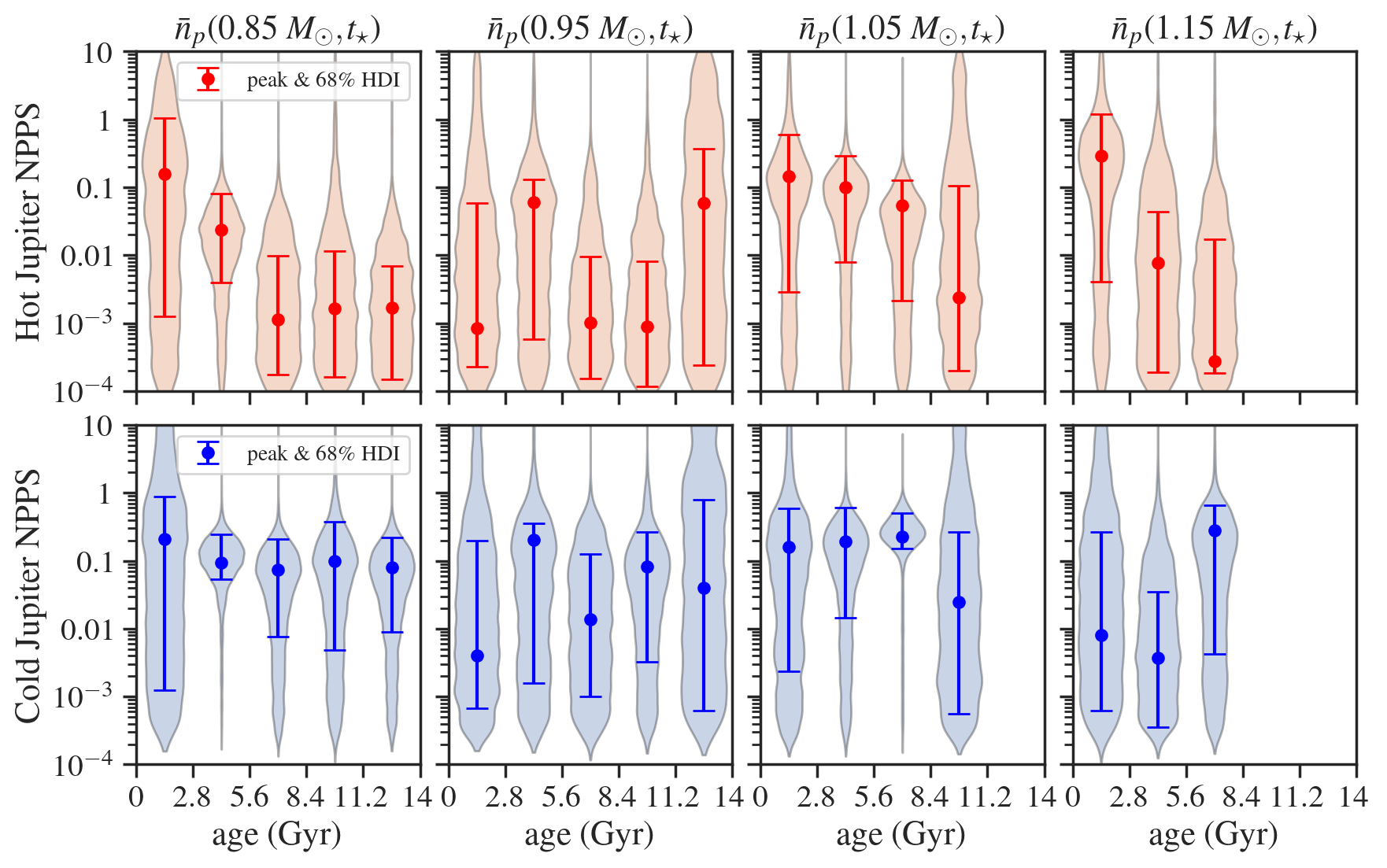}
    \caption{
    The posterior PDFs for the NPPS as a function of age derived using the histogram model in Section~\ref{subsec:histogram}. Unlike in the right panels of Figure \ref{fig:NPPS_nonpara}, here we show the NPPS evaluated at given stellar masses shown at the top of each column, after marginalizing over the metallicity distribution.
    }
    \label{fig:NPPS_mass_age}
\end{figure}

In this section, we apply the framework described in Section \ref{sec:framework} to the Sun-like sample defined in Section~\ref{sec:sample} and infer the occurrence rate of hot Jupiters (HJs) and cold Jupiter (CJs), taking into account the dependencies on stellar and planetary properties.
For the stellar properties, we consider metallicity, mass, and age, i.e., $\bm{z}=(\mathrm{[Fe/H]}, M_\star,t_\star)$, and ignore possible dependencies on other parameters including the Galactic position and velocity. 
As we discussed in Section~\ref{ssec:simple_analysis}, it is essential to consider all these variables jointly so that the dependencies on mass and/or metallicity do not mimic the age dependence through the correlations between these parameters. 
For the planet properties, we consider the logarithms of the minimum mass and orbital period that are well determined from RVs: $\bm{x}=(\log M\sin{i}, \log P)$.

We assume the following functional form for the NPPS function $f(\bm{x}|\bm{\gamma},\bm{z})$: 
\begin{eqnarray}
\label{eq:f_model}
    f(\bm{x}|\bm{\gamma},\bm{z}) &=& \frac{\partial^2 \bar{n}_p} 
    {\partial({\log{M\sin{i}}}) \partial(\log{P})} \nonumber\\
    &=& \left\{
    \begin{array}{lc}
        \mathcal{G}(\bm{z},\bm{\gamma}_s) \left({M\sin{i}}/{M_{\rm Jup}}\right)^m \left({P}/{\rm day}\right)^p,  & \mathrm{(for\;HJ)} \\
        \mathcal{G}(\bm{z},\bm{\gamma}_s) \left({M\sin{i}}/{M_{\rm Jup}}\right)^m \left({P}/{\rm year}\right)^p,  & \mathrm{(for\;CJ)} 
    \end{array}
    \right.\nonumber\\
\end{eqnarray}
where $\bm{\gamma}=\{\bm{\gamma}_s, m, p\}$ are the parameters that specify $f$.
Here, $\mathcal{G}(\bm{z},\bm{\gamma}_s)$ is a shape function that describes the dependency on stellar properties, and it includes a normalization factor for the NPPS.
We follow previous works \citep[e.g.,][]{Howard+2010, Petigura+2018} to model the dependence on the planet properties as a double power law with indices $m$ and $p$.
Although recent studies report a possible deviation from a single power law function at long orbital periods corresponding to a few au \citep{Fernandes+2019, Fulton+2021}, a single power law still remains a reasonable model for the narrow range of our CJs' period (1--10~yr). 
By Equation~(\ref{eq:f_model}), we implicitly assume that the mass/period dependences are separable, and are common for all stars with different $\bm{z}$.

For evaluating the detection efficiency for each star, $\eta_j(\bm{x})$, we divide the parameter space of planetary properties $\bm{x}$ into cells with dimensions of $\Delta(\log{P}) \times \Delta(\log{M\sin{i}})\sim0.14\;{\rm dex} \times 0.13\;{\rm dex}$, and computed the value of $\eta_j$ for each of these cells using the results of injection-recovery simulations \citep{Rosenthal+2021}.
The efficiency averaged over all the sample stars is shown in Figure \ref{fig:planets_DE}.

\subsection{Histogram Model\label{subsec:histogram}}

\begin{deluxetable}{cc|cc}[ht!]
\tablecaption{Priors and Posteriors of Parameters in the Histogram Inference \label{tab:histogram_table}}
\tablehead{
\multirow{2}{*}{Parameters} & \multirow{2}{*}{Prior} & \multicolumn{2}{c}{Posterior summary} \\
 &  &\multicolumn{1}{c}{Hot Jupiter} & \multicolumn{1}{c}{Cold Jupiter} 
}
\startdata
\multicolumn{3}{l}{\it Parameters for the NPPS function $f(\bm{x}|\bm{z})$}\\
$\{\log f_k\}$ & $\mathcal{U}(-4,1)$ & \multicolumn{2}{c}{see Figure~\ref{fig:NPPS_nonpara}}\\
$m$ & $\mathcal{U}(-5,5)$ & $-0.68^{+0.32}_{-0.33}$ & $-0.18^{+0.15}_{-0.15}$\\
$p$ & $\mathcal{U}(-5,5)$ & $-0.34^{+0.45}_{-0.46}$ & $0.51^{+0.25}_{-0.25}$\\ 
\multicolumn{3}{l}{\it Parameters for the stellar distribution $p_\star(\bm{z})$}\\ 
$\{\alpha_k\}$ & Eq.~(\ref{eq:prior_alpha})  & \multicolumn{2}{c}{see Figure~\ref{fig:occ_nonpara}}
\enddata
\tablecomments{The values shown are the medians and 68\% equal-tail intervals of the marginal posterior distribution.
}
\end{deluxetable}
Throughout this paper, the stellar distribution $p_\star(\bm{z})$ is modeled as a histogram, as shown in Equation~(\ref{eq:star_dist}). 
In this subsection, we model $\mathcal{G}(\bm{z},\bm{\gamma}_s)$ as a histogram too, using the common bins in the $\bm{z}$ space as adopted for $p_\star(\bm{z})$.
This is meant to serve as an analysis with minimal assumptions on the $\bm{z}$ dependence of the NPPS, which guides the interpretation of the subsequent results from other more constrained parametric models. 
We set up $4 \times 4 \times 5$ bins for [Fe/H], mass, and age spanning $[-0.4,0.4]$, $[0.8,1.2]$, and $[0,14]$, respectively, resulting in the total bin number of $K=80$. 
The bin widths are determined to be comparable to the typical uncertainty of these parameters (see also Section~\ref{ssec:conclusion}); we do not expect that the data provide useful information on finer parameter dependencies.
Then we represent the expected NPPS values in each bin (assumed to be constant within the bin) by $\bm{\gamma}_s=\{f_1,f_2,...,f_K\}$. 
Consequently, we have 
\begin{align}
    \mathcal{G}(\bm{z},\bm{\gamma}_s) = \sum^{K}_{k=1} {f_k \over \Delta_m \Delta_p} \bm{1}_k(\bm{z}),
\end{align}
where $\Delta_m \Delta_p$ denotes the integral of the power law part of Equation~(\ref{eq:f_model}) in the HJ/CJ domain, and depends both on $m$ and $p$. 
This factor makes $f_k$ to be the number of planets in the domain per star with parameters $\bm{z}$; see Equation~(\ref{eq:np_z}).

The prior $p(\bm{\alpha},\bm{\gamma})$ is assumed to be separable as $p(\bm{\alpha},\bm{\gamma})=p(\bm{\alpha})\,p(\bm{\gamma})$. 
For $\bm{\alpha}$ that describes the stellar distribution $p_\star(\bm{z}|\bm{\alpha})$, we assume
\begin{eqnarray}\label{eq:prior_alpha}
    p(\bm{\alpha}) &\propto& \delta\left(\sum^{K}_{k}\exp(\alpha_k)\Delta_k-1\right) \nonumber\\
    &\times& \left[
    \prod_{k \notin \mathcal{K}} \mathcal{U}(\alpha_k;-5,\alpha_{\rm max})\right] \cdot 
    \left[\prod_{k \in \mathcal{K}} \delta(\alpha_k + 10)\right],
\end{eqnarray}
where $\delta(x)$ denotes the Dirac delta function, and $\mathcal{U}(x;a,b)$ denotes the uniform distribution for $x$ between $a$ and $b$. 
The first term on the right side comes from the normalization requirement of the PDF. 
In the second line, we assign uniform priors on most $\alpha_k$, but assign essentially zero probability density $\alpha_k=-10$ at bins in the set $\mathcal{K}$ that satisfies
\begin{eqnarray}\label{eq:prior2}
    t_\star\;{\rm (Gyr)} > -30(M_\star/M_\odot -1.25)+5.
\end{eqnarray}
This represents our prior knowledge that stars do not exist in these regions.
The value of $\alpha_{\rm max}$ is chosen so that the normalized $p_\star(\bm{z}|\bm{\alpha})$ is positive (i.e., $\exp(\alpha_k)\Delta_k$ never exceeds unity for any $k$).
For the parameters $\bm{\gamma}$ in the NPPS function $f(\bm{x}|\bm{\gamma}, \bm{z})$, we assume independent and uniform priors:
\begin{eqnarray}\label{eq:prior_gamma}
    p(\bm{\gamma}) = \left[\prod^{K}_{k=1}\mathcal{U}(\log{f_k}; -4, 1)\right]\,\mathcal{U}(m; -5, 5)\,\mathcal{U}(p; -5, 5) \nonumber \\
\end{eqnarray}

We obtain 10,000 samples from the posterior distribution in Equation (\ref{eq:summary}) and confirm $\hat{R}<1.01$ for all the parameters.
Figure \ref{fig:occ_nonpara}, Figure~\ref{fig:NPPS_nonpara}, Figure~\ref{fig:NPPS_mass_age}, and Table~\ref{tab:histogram_table} present a summary of the result for the joint inference of the stellar parameter distribution $p_\star(\bm{z})$ and the NPPS function $f(\bm{x}|\bm{\gamma},\bm{z})$.
Figure~\ref{fig:occ_nonpara} shows the stellar parameter distribution $p_\star(\bm{z})$ inferred from the whole data.
The left panel displays the joint distribution:
\begin{align}
    p(\bm{z}|\bm{D}, \bm{H}) = \int p_\star(\bm{z}|\bm{\alpha})\,p(\bm{\alpha},\bm{\gamma}|\bm{D}, \bm{H})\,d\bm{\alpha}d\bm{\gamma}.
\end{align}
The right panels show the marginalized distributions of this joint PDF $p(\bm{z}|\bm{D}, \bm{H})$ for [Fe/H], mass, and age. 
Figure \ref{fig:NPPS_nonpara} shows the inferred NPPS functions for HJs and CJs. 
What is shown in the right panels is essentially Equation~(\ref{eq:np_z}) conditioned on the data, but for 1D visualization as a function of each of the mass, metallicity, and age (which we denote by $\hat{z}$), it has been marginalized over the other two stellar parameters ($\bm{z}_{\backslash \hat{z}}$). 
Namely, shown here by the violin plots are the distributions of
\begin{align}
\label{eq:violin}
    \int \overline{n}_{\bm{x}\in\mathcal{P}}(\bm{\gamma}, \bm{z})\,p_\star(\bm{z}_{\backslash \hat{z}}|\bm{\alpha}, \hat{z})\,d\bm{z}_{\backslash \hat{z}}
\end{align}
(see also Equation~(\ref{eq:np_z})) computed for the samples of $(\bm{\alpha}, \bm{\gamma})$ drawn from $p(\bm{\alpha}, \bm{\gamma}|\bm{D}, \bm{H})$.
From these plots, we see that the NPPS of both HJs and CJs is likely higher for more metal-rich stars. 
This is a known trend, but we recover this even considering the correlations with other stellar parameters. 
Furthermore, the NPPS of HJs shows a decreasing trend with increasing age, to a similar degree to the metallicity dependence. We further investigate these trends with the parametric models below. 
The left panel of Figure \ref{fig:NPPS_nonpara} shows the posterior PDFs for the NPPS of HJs and CJs in the Sun-like sample, averaged over the stellar parameters: the histograms show the distributions of Equation~(\ref{eq:np}), $\overline{n}_{\bm{x}\in\mathcal{P}, \bm{z}\sim p_\star(\bm{z})} (\bm{\alpha}, \bm{\gamma})$, computed for the samples of $(\bm{\alpha}, \bm{\gamma})$ drawn from $p(\bm{\alpha}, \bm{\gamma}|D, H)$.
The posterior median and equal-tail 68\% intervals are $4.0^{+1.9}_{-1.2}\,\%$ for HJs (orange) and $18.0^{+5.1}_{-3.3}\,\%$ for CJs (blue), respectively.
They are both consistent with the values reported by \citet{Zhu+2022}, who analyzed a similar subset of Sun-like stars from the CLS catalog.
In Figure~\ref{fig:NPPS_mass_age}, we show the posterior PDFs for the NPPS as a function of both mass and age, now marginalizing only over the metallicity. Although the results for separate stellar mass bins are more uncertain, the NPPS of HJs still shows a decreasing trend with increasing age at each stellar mass.
This confirms that the age dependence seen in Figure~\ref{fig:NPPS_nonpara} is not solely due to the mass dependence (cf. Section~\ref{ssec:simple_analysis}); even if we control the stellar mass, the NPPS of HJs tends to be higher at younger ages.

\begin{figure*}[t!]
    \centering
    \includegraphics[bb=0 0 1090 445,scale=0.46]{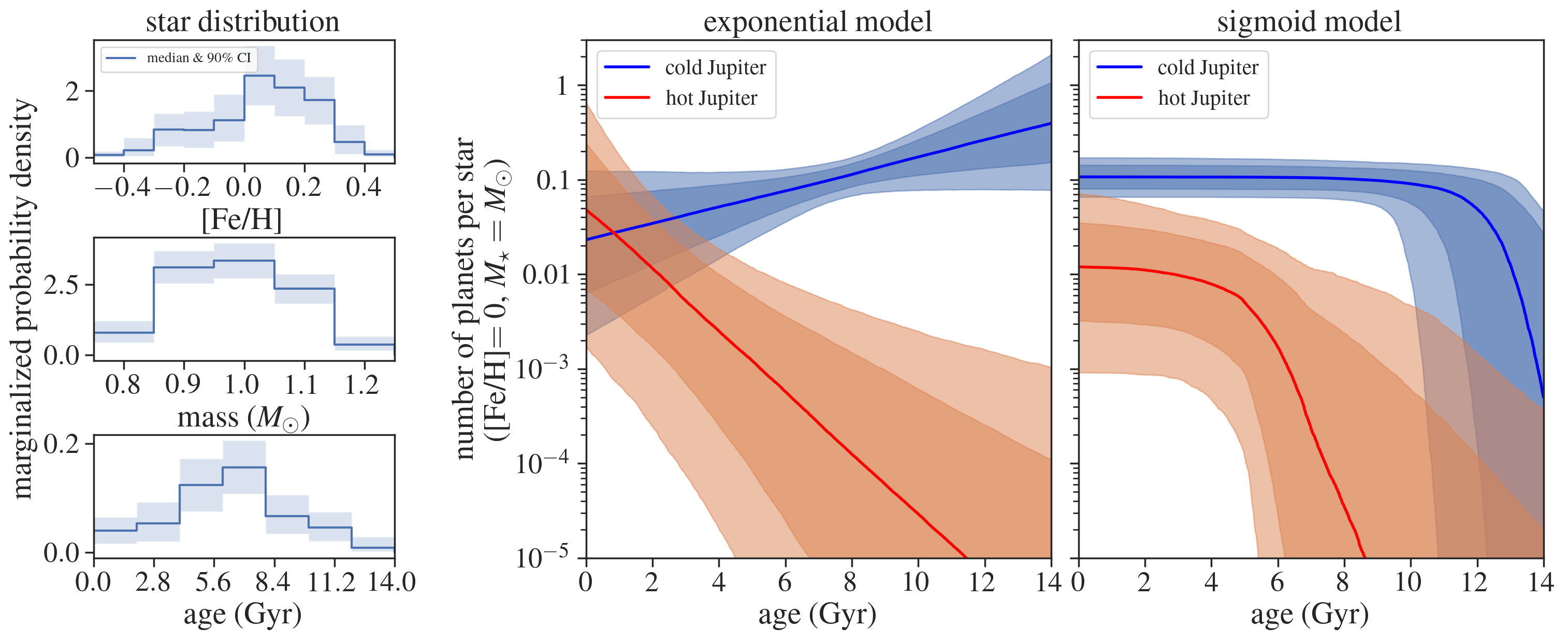}
    \caption{
    ({\bf Left}) The panels show the marginalized posterior PDFs for [Fe/H], mass, and age (top to bottom).
    The solid lines and translucent regions represent the medians and 90\% credible intervals (CIs) in each bin.
    ({\bf Right}) The inferred NPPS are functions of age for hot Jupiters and cold Jupiters, adopting the exponential model (left panel) and the sigmoid model (right panel). 
    This is Equation~(\ref{eq:np_z}) evaluated at the solar mass and metallicity.
    In each panel, solid lines present median values of the posterior predictions at ages and dark and light-shaded regions means $68\%$ and $90\%$ credible intervals of the posterior predictions, respectively.   
    }
    \label{fig:summary_parametric}
\end{figure*}

\subsection{Other Parametric Models}\label{subsec:parametric}
\begin{deluxetable*}{cc|ccccc}[ht!]
\tablecaption{Priors and Posteriors of Parameters in the Parametric Inference \label{tab:parametric}}
\tablehead{
\multirow{2}{*}{Parameters} & \multirow{2}{*}{Prior} & \multicolumn{2}{c}{Exponential} & \multicolumn{2}{c}{Sigmoid} \\
 &  &\multicolumn{1}{c}{Hot Jupiter} & \multicolumn{1}{c}{Cold Jupiter} & \multicolumn{1}{c}{Hot Jupiter} & \multicolumn{1}{c}{Cold Jupiter}
}
\startdata
\multicolumn{3}{l}{\it Parameters for the NPPS function $f(\bm{x}|\bm{z})$}\\
$\log{f_0}$     & $\mathcal{U}(-5,0)$       & $-1.65^{+0.76}_{-0.90}$  & $-2.17^{+0.48}_{-0.60}$ & $-2.25^{+0.57}_{-0.67}$ & $-1.51^{+0.20}_{-0.20}$  \\
$m$             & $\mathcal{U}(-5,5)$       & $-0.75^{+0.35}_{-0.38}$  & $-0.16^{+0.16}_{-0.16}$ & $-0.75^{+0.35}_{-0.37}$ & $-0.18^{+0.15}_{-0.15}$  \\
$p$             & $\mathcal{U}(-5,5)$       & $0.34^{+0.53}_{-0.51}$   & $ 0.66^{+0.26}_{-0.25}$  & $ 0.33^{+0.53}_{-0.52}$ & $0.67^{+0.27}_{-0.27}$  \\
$\beta$         & $\mathcal{U}(-20,20)$     & $4.62^{+3.91}_{-2.37}$   & $ 1.86^{+0.77}_{-0.71}$ & $ 3.36^{+2.38}_{-1.78}$ & $ 1.33^{+0.65}_{-0.63}$  \\
$\kappa$        & $\mathcal{U}(-30,30)$     & $12.52^{+8.88}_{-6.87}$  & $ 1.46^{+2.60}_{-2.60}$ & $ 8.98^{+6.61}_{-5.90}$ & $ 2.13^{+2.35}_{-2.24}$  \\
$\epsilon$      & $\mathcal{U}(-5,5)$       & $-0.74^{+0.34}_{-0.54}$  & $ 0.20^{+0.15}_{-0.14}$ & -                       & -                        \\
$\log{\lambda}$ & $\mathcal{U}(-0.5,1.5)$   & -                        & -                       & $ 0.35^{+0.77}_{-0.48}$ & $ 0.51^{+0.66}_{-0.66}$  \\
$C_{\rm age}$   & $\mathcal{U}(0,14)$       & -                        & -                       & $ 5.13^{+1.80}_{-2.34}$ & $ 11.94^{+1.43}_{-1.77}$ \\
\multicolumn{3}{l}{\it Parameters for the stellar distribution $p_\star(\bm{z})$}\\
$\{\alpha_k\}$ & Eq.~(\ref{eq:prior_alpha})  & \multicolumn{2}{c}{see Figure~\ref{fig:summary_parametric} left} & \multicolumn{2}{c}{not shown but similar to Figure~\ref{fig:summary_parametric}}
\enddata
\tablecomments{The values shown are the medians and 68\% equal-tail intervals of the marginal posterior distribution. 
$f_0$: normalization factor, $m$: power law index for planet mass, $p$: power law index for orbital period, $\beta$: power law index for metallicity, $\kappa$: power law index for stellar mass, $\epsilon$: index for the exponential form, $\lambda$: decrease timescale for the sigmoid function, $C_{\rm age}$: cut-off age for the sigmoid function.
See Equation (\ref{eq:parametric}), (\ref{eq:exponential}), and (\ref{eq:sigmoid}).
}
\end{deluxetable*}

In this section, we model $\mathcal{G}(\bm{z},\bm{\gamma}_s)$ as a parametric function of the form:
\begin{eqnarray}\label{eq:parametric}
    \mathcal{G}(\bm{z},\bm{\gamma}_s) = f_0 10^{\beta \mathrm{[Fe/H]}} (M_\star/M_\odot)^{\kappa} \mathcal{G}^{\prime}(t_\star,\bm{\gamma}^{\prime}),
\end{eqnarray}
where $f_0$ is a normalization factor for the NPPS, and the function $\mathcal{G}^{\prime}(t_\star,\bm{\gamma}^{\prime})$ that accounts for the age dependence will be specified below. 
The metallicity and mass dependencies are modeled as power-law functions with indices $\beta$ and $\kappa$, respectively, following the prior studies \citep[e.g.,][]{Johnson+2010}.
We set uniform priors for $\beta$ and $\kappa$ within $[-20,20]$ and $[-30,30]$, respectively, and a log-uniform prior for $f_0$ within $[10^{-5},1]$; see Table~\ref{tab:parametric}.

We continue to model the stellar distribution as a histogram as in Section \ref{subsec:histogram}, but given the fewer number of free parameters in $\mathcal{G}(\bm{z}, \bm{\gamma}_s)$, we increase the resolution. 
We set up 10 bins for [Fe/H] spanning $[-0.5,0.5]$ at 0.1 dex intervals, 5 bins for mass spanning $[0.75,1.25]\; M_\odot$ at $0.1\,M_\odot$ intervals, 7 bins for ages spanning $[0,14]$ Gyr at 2 Gyr intervals, which results in the total number of bins of $K=10\times5\times7=350$.
The prior PDF for $\bm{\alpha}$ is again given by Equation~(\ref{eq:prior_alpha}).

\subsubsection{Exponential Function for Age}
First, we assume an exponential function for the age dependence:
\begin{equation}\label{eq:exponential}
    \mathcal{G}^{\prime}(t_\star,\bm{\gamma}^{\prime})=e^{\epsilon (t_\star/\mathrm{Gyr})}.
\end{equation}
This form is not necessarily motivated physically but will be useful to characterize the timescale of a monotonic change in the NPPS, which is given by $1/\epsilon$.
We set a uniform prior for $\epsilon$ within $[-5,5]$, allowing for the NPPS to increase, decrease, or remain constant as a function of age. 

We obtain 10,000 samples from the posterior distribution in Equation (\ref{eq:summary}), where $\bm{\gamma}=\{m, p, \log f_0, \beta, \kappa, \epsilon \}$. 
The posterior constraints on the parameters $\bm{\gamma}$ are summarized in Table \ref{tab:parametric}.
The corner plot for $\bm{\gamma}$ is shown in Figure~\ref{fig:MCMC_exp} in Appendix.
The left panel of Figure \ref{fig:summary_parametric} presents the inferred stellar distribution, and the right panel shows the inferred NPPS as a function of age for HJs (red) and CJs (blue). 
Here we show the median and 68\%/90\% equal-tail intervals of $f(\bm{x}|\bm{\gamma}, \bm{z}=\{t_\star, \mathrm{[Fe/H]}=0, M_\star=M_\odot\}$ integrated over $\bm{x}$ and evaluated for the posterior samples of $\bm{\gamma}$.
We obtain $\epsilon_{\rm HJ}=-0.74_{-0.98}^{+0.54}$ for HJs and $\epsilon_{\rm CJ}=0.20_{-0.23}^{+0.28}$ for CJs (both 90\% equal-tail credible intervals) and find that the NPPS of HJs likely decreases with increasing age, while that for CJs is consistent with a constant value, in agreement with the result of the histogram inference.
Assuming that $\epsilon_{\rm HJ}<0$, which is satisfied by more than 99\% of the posterior samples, the timescale for the decrease of HJ NPPS is found to be $-1/\epsilon_{\rm HJ}=1.35^{+1.09}_{-0.57}\,\mathrm{Gyr}$ (68\% credible interval).

We find positive metallicity dependency $\beta$ for both HJ and CJ, which is consistent with prior studies \citep[e.g.,][]{Petigura+2018}.
As expected from the histogram result, we do not find very strong evidence for the stellar mass dependence (i.e., non-zero $\kappa$) in our Sun-like star sample.
This is possibly due to the narrower mass range than investigated by \citet{Johnson+2010}; or such dependence may not really exist \citep[e.g.,][]{2019AJ....158..141Z}.
Because we model these dependencies simultaneously with the age dependence, the analysis suggests that the metallicity dependence of giant planets' NPPS found in previous works is unlikely to be an artifact caused by the dependence on age and its correlation with metallicity.

To check on the possible prior dependence, we repeated a similar analysis by re-parametrizing $\epsilon=\tan\theta$ and by assigning a uniform prior between $-\pi/2$ and $\pi/2$ for $\theta$. This choice results in the prior PDF for $\epsilon$ that is more narrowly peaked around zero than in the model above, putting more weight on the solutions without age dependence. The result was found to be consistent with the above analysis adopting a uniform prior for $\epsilon$.

\subsubsection{Sigmoid Function for Age}

Next, we assume a sigmoid function of the form 
\begin{equation}\label{eq:sigmoid}
    \mathcal{G}^{\prime}(t_\star,\bm{\gamma}^{\prime})=\left[1+e^{\lambda(t_\star/\mathrm{Gyr}-C_{\rm age})}\right]^{-1}.
\end{equation}
This function remains almost constant up to around the cut-off age $C_{\rm age}$, and decreases over a timescale characterized by the parameter $\lambda\;(>0)$.
We assume a log-uniform prior for $\lambda$ within $[10^{-0.5},10^{1.5}]$ and a uniform prior for the cut-off age $C_{\rm age}$ between 0 and 14. 
With the two free parameters, this model has a larger flexibility than the exponential model. 
When $\lambda$ is large, $\mathcal{G}^{\prime}$ approaches a step function, while when it is small $\mathcal{G}^{\prime}$ behaves in a similar manner to the exponential function as adopted in the previous subsection.

From Equation~(\ref{eq:summary}), we obtain 10,000 posterior samples, where $\bm{\gamma}=\{m, p, \log f_0, \beta, \kappa,  \lambda, C_{\rm age} \}$. 
The result of this inference is summarized in Table~\ref{tab:parametric}, Figure~\ref{fig:summary_parametric}, and Figure~\ref{fig:MCMC_sig}.
The right panel of Figure \ref{fig:summary_parametric} presents the inferred NPPS for the sigmoid model. 
Although the parameters $\lambda$ and $C_{\rm age}$ are strongly degenerate for HJs, the inferred NPPS functions show a similar tendency as seen for the exponential model: the NPPS of HJs decreases with age, while that of CJs does not, at least up to $C_{\rm age}\sim 12\,\mathrm{Gyr}$. 
While the NPPS of HJs at ages less than a few Gyr appears different from the exponential case by the construction of the sigmoid model, the 68\% regions from the two models overlap.
Similar to the exponential model, the sigmoid model also suggests that the age dependences (i.e., shapes of $\mathcal{G}'(t_\star, \gamma')$) for HJs and CJs are likely different; see also the corner plot (Figure \ref{fig:MCMC_sig}) in Appendix.

\subsection{Conclusions}\label{ssec:conclusion}

Both the histogram analysis (Section~\ref{subsec:histogram}) and the other parametric analysis adopting two different functional forms (Section~\ref{subsec:parametric}) consistently suggest that the NPPS of HJs decreases with increasing age. 
The parametric models give tighter constraints on the NPPS than the histogram one, presumably because of the additional assumption that the NPPS changes smoothly as a function of age. 
The consistency of the results based on two different parametric models suggests that the inferred age dependence is not too sensitive to how this smoothness is implemented in the NPPS function. 

The NPPS of HJs decreases in the latter half of the main-sequence lifetime, while the behavior in the first half is not well constrained partly due to the small number of stars searched (see Figure~\ref{fig:summary_parametric} left); the lack of young stars may be due to the selection of the Doppler survey against active stars that are not suited for precise RV measurements. 
Although the uncertainty in the inferred NPPS is still large due to the limited number of HJs, the result suggests that the change in the NPPS around older main-sequence stars could well be quite rapid: both of the parametric models suggest that the NPPS decreases by an order of magnitude over a few Gyrs. 

In contrast, the NPPS of CJs does not show strong evidence for age dependency during the main sequence. 
Given that we do not know of physical processes that change their occurrence rate during the main sequence, this absence of age dependence supports that our inference framework is working properly. 
For evolved stars, on the other hand, the sigmoid model may hint at the decrease of NPPS; we are not fully sure if this is a real trend, given that this may partly be driven by the construction of the sigmoid model, and that the NPPS is not well constrained due to the lack of such old stars. 
The slight inconsistency in the inferred NPPS of CJs around the oldest stars based on the two parametric models implies that the NPPS of CJs may have a more complicated dependence on age than we assume here and that the available data may already provide such information. It is beyond the scope of the present work to perform a more detailed study of the NPPS evolution for CJs.

In this work, we adopted the MIST models for the stellar isochrone fitting, and have not examined the dependence of the results on the adopted stellar model.
\citet{Tayar+2022} demonstrated that the model-dependent offsets are typically $\sim5\%$ in mass and $\sim20\%$ in age for main-sequence and sub-giant stars.
Considering the current precision of the inference, these model-dependent offsets do not significantly affect the above conclusions but may become more important in future analyses using larger samples. 
Because our analysis is based on the homogeneous analysis of stars using the same stellar model and the observing data with similar qualities, the conclusions based on relative comparisons, such as the NPPS difference between younger and older stars, and the difference between HJs and CJs, are more robust against the model-dependent systematics than those based on the absolute age (e.g., at what age the NPPS starts decreasing).

\section{Discussion \label{sec:discussion}}
\subsection{Implications for the Tidal Evolution of HJs}

Tidal orbital decay appears to be the most plausible explanation for the decrease of HJ's NPPS around older stars (Section~\ref{sec:inference}).
In principle, the NPPS as a function of age, as obtained in this work, enables the joint inference of the mass-period distribution of HJs before tidal evolution as well as the efficiency of tidal dissipation inside the star. 
Because the NPPS analysis uses the information for all the stars that have been surveyed, such an analysis can consider both HJs that currently survive and that have already been engulfed in a consistent manner, which is otherwise difficult. 
It is even possible, at least in principle, to include the dependence of the dissipation efficiency on the star's mass and age, tidal forcing frequency, and planet's mass. 
The information on the ``initial" mass--period distribution may be useful to distinguish the origin scenarios of HJs. 
The empirical constraint on the tidal dissipation efficiency will be useful for improving the theory of tidal dissipation, and also for better understanding the orbital evolution of stellar binaries in general. 

We leave this ``full" analysis for future work, given that the NPPS function based on the CLS sample still has a large uncertainty.
Instead, here we give a simple order-of-magnitude estimate for the tidal dissipation efficiency based on our inference, assuming that the decrease of the HJ's NPPS is real and solely due to the tidal orbital decay. 
We characterize the efficiency of tidal dissipation by the modified stellar tidal quality factor, $Q^\prime_\star=3Q_\star/2k_2$, where $Q_\star$ represents the ratio of the maximum energy stored in the tide to the energy dissipated in one cycle, and $k_2$ denotes the tidal Love number (equals 3/2 for a homogeneous sphere). 
Highly dissipative stars have small $Q^\prime_\star$ values, and the $Q^\prime_\star$ value, in general, depends on tidal forcing frequency and amplitude as well as on the internal structure of the star \citep[e.g.,][]{Barker2020}. 
If $Q^\prime_\star$ is assumed to be constant, the inspiral time for a planet with orbital period $P$ is analytically given by \citep[e.g.,][]{Ogilvie+2014}:
\begin{eqnarray}\label{eq:tin}
    t_{\rm in} 
    &\simeq& 2.3\,\mathrm{Gyr}\left(Q^\prime_\star \over 10^6\right) \left(q \over 10^{-3}\right)^{-1} \left(\frac{\bar{\rho}_\star}{\bar{\rho}_\odot}\right)^{\frac{5}{3}} \left( \frac{P}{3\,{\rm days}}\right)^{\frac{13}{3}}\nonumber,\\
\end{eqnarray}
where $q$ is the planet-to-star mass ratio, and $\bar{\rho}_\star$ and $\bar{\rho}_\odot$ are the mean densities of the host star and the Sun, respectively. 
The values of $q$ and $P$ in the above equation correspond to typical HJs in our sample (see Figure~\ref{fig:planets_DE}), which should be considered as planets that have not yet been engulfed by the stars and mostly preserve their initial orbits (because the change in the orbit is small except for the last moment of tidal engulfment). 
The results of the parametric inferences in Figure~\ref{fig:summary_parametric} imply that they will be lost during the first half of the main sequence, roughly within $\sim 6\,\mathrm{Gyr}$. 
Thus Equation~(\ref{eq:tin}) implies $Q^\prime_\star \sim 10^6$ for a typical HJ orbiting a typical Sun-like star in our sample. 
We note that the corresponding inspiral time in Equation~(\ref{eq:tin}) is also comparable to the $\sim 1\,\mathrm{Gyr}$ timescale inferred from our exponential model.
This estimate of the tidal quality factor is essentially similar to $\log_{10}{Q^\prime_\star}<6.5^{+0.5}_{-0.6}$ given by \citet{Hamer+2019} who required $t_{\rm in}$ to be shorter than the main-sequence lifetime of Sun-like stars. 
The estimate is also comparable to the theoretical estimate assuming the efficient non-linear dissipation of internal gravity waves for the tidal forcing frequency of $\approx 3\,\mathrm{days}/2=1.5\,\mathrm{days}$ \citep[see Equation 54 of][]{Barker2020}.\footnote{\citet{Barker2020} derived $Q^\prime_\star$ scales as $P_{\rm tide}^{8/3}$ with $P_{\rm tide}$ being tidal forcing frequency. 
If we modify Equation~\ref{eq:tin} considering this dependence and assuming $P_{\rm tide}=P/2$, and apply the same argument, we estimate $Q^\prime_\star$ at $P_{\rm tide}=0.5\,\mathrm{days}$ to be $\sim 10^5$ instead. This also agrees with his Equation~54.}

\subsection{Implications for Past and Future Survey Results}

In this work, we focused on $382$ Sun-like stars from the CLS survey, which includes nine HJs. Due to this small number, the timescale for the decline of the NPPS is still consistent with a wide range of values. Our framework can be extended to transit surveys, and will also be useful to investigate HJ samples that are more than an order of magnitude larger \citep[e.g.,][]{2023ApJS..265....1Y}. Such applications would provide more precise age dependence of the NPPS, and might also reveal the age dependence of the mass--period distribution of HJs: in this work, we ignored possible stellar parameter dependence of the shape of $f(\bm{x}|\bm{\gamma}, \bm{z})$ (i.e., $m$ and $p$ in Equation~\ref{eq:f_model} do not depend on stellar parameters) but this assumption may be relaxed with a larger sample.

Our result suggests that the age dependence of the HJ occurrence could be as important as that on metallicity. 
Then it is crucial to consider the age distribution of the sample stars as well in interpreting the results of different surveys. 
A general tendency would be that surveys targeting a larger fraction of older, brighter, more evolved Sun-like stars tend to give lower HJ occurrence rates, all else being equal.
The age dependency might turn out to be the key to explaining marginally different survey results as mentioned in Section~\ref{sec:intro} in a unified manner.
Such consideration will be even more important as the \textit{Nancy Grace Roman Space Telescope} will uncover a large number of HJs orbiting stars across the Milky Way, including those in the Galactic Center \citep{Montet+2017, Miyazaki+2021, Wilson+2023} where we expect a large difference in the age distribution from the solar neighborhood. It is also possible for our framework to take into account the NPPS dependence on stellar properties other than mass, age, and metallicity, such as the position within the Galaxy. 

\section{Summary \label{sec:summary}}

In this paper, we developed a Bayesian framework to infer the number of planets per star (NPPS) as a function of both planetary and stellar properties, which must involve the simultaneous inference of the stellar parameter distribution in the surveyed stars (Section~\ref{sec:framework}).
Our framework can handle large and heterogeneous uncertainties in the degenerate stellar parameters by incorporating the information on the entire likelihood function for the stellar parameters.
We applied the framework to hot Jupiters (HJs) and cold Jupiters (CJs) orbiting Sun-like stars from the California Legacy Survey (Section~\ref{sec:sample})
to derive their NPPSs as a function of the stellar mass, age, and metallicity, which were estimated 
by fitting the isochrone models to their measured spectroscopic parameters, Gaia DR3 parallaxes, and 2MASS $K_{\rm s}$-band magnitudes.
We found evidence that the NPPS of HJs decreases in the latter half of the main-sequence lifetime of their Sun-like host stars over the timescale of several Gyr, while the NPPS of CJs does not show the same trend (Section~\ref{sec:inference}).
The result implies that the modified stellar tidal quality factor $Q^\prime_\star$ is on the order of $10^6$ for a typical HJ in the sample on a $\approx 3\,\mathrm{day}$ orbit (Section~\ref{sec:discussion}).

\section*{Acknowledgments}
The authors thank Luke Bouma, Morgan MacLeod, Kaloyan Penev, and Josh Winn for their helpful comments on the manuscript. 
We also thank the anonymous reviewers for constructive comments that helped us improve the manuscript.
We sincerely appreciate all the efforts and contributions to constructing the California Legacy Survey and making their data and codes publicly available.
This work has made use of data from the European Space Agency (ESA) mission {\it Gaia} (\url{https://www.cosmos.esa.int/gaia}), processed by the {\it Gaia} Data Processing and Analysis Consortium (DPAC, \url{https://www.cosmos.esa.int/web/gaia/dpac/consortium}). 
Funding for the DPAC has been provided by national institutions, in particular, the institutions participating in the {\it Gaia} Multilateral Agreement.
This publication makes use of data products from the Two Micron All Sky Survey, which is a joint project of the University of Massachusetts and the Infrared Processing and Analysis Center/California Institute of Technology, funded by the National Aeronautics and Space Administration and the National Science Foundation.
\software{
{\tt corner} \citep{corner},
{\tt JAX} \citep{Bradbury+2021},
{\tt Numpyro} \citep{phan2019composable,bingham2019pyro},
{\tt jaxstar} \citep{Masuda2022, jaxstar}
}

\appendix
\section{Special Cases of the Formulation in Section~\ref{sec:framework}\label{sec:various_case}}

Here we describe special cases of our general formulation presented in Section~\ref{sec:framework}, and show that the results in previous works are recovered.

\subsection{Ignoring Dependency on Stellar Properties}

When the dependency of the NPPS function on stellar properties $\bm{z}$ is ignored, Equation (\ref{eq:DjHj}) reduces to: 
\begin{eqnarray}
    &&p(D_j,H_j|\bm{\alpha},\bm{\gamma}) =  p(H_j|\bm{\gamma})p(D_j|\bm{\alpha}) \nonumber\\
    &=&\left(\prod_{\bm{x}_j\in H_j} \widetilde{f}_j(\bm{x}_j|\bm{\gamma})\right)\exp{\left[-\int \widetilde{f}_j(\bm{x}|\bm{\gamma})d\bm{x} \right]} \nonumber \\ 
    && \times \int p(D_j|\bm{z}_j) p_\star(\bm{z}_j|\bm{\alpha}) d\bm{z}_j.
\end{eqnarray}
Therefore
\begin{align}
    \mathcal{L}(\bm{\alpha}, \bm{\gamma}) = \mathcal{L}(\bm{\gamma})\,\mathcal{L}(\bm{\alpha}),
\end{align}
where
\begin{eqnarray}\label{eq:no_z_NPPS}
    \mathcal{L}(\bm{\gamma}) &=& \prod^{N_\star}_j \left(\prod_{\bm{x}_j\in H_j} \widetilde{f}_j(\bm{x}_j|\bm{\gamma})\right) \exp{\left[-\int \widetilde{f}_j(\bm{x}|\bm{\gamma})d\bm{x} \right]}, \nonumber \\ \\
    \mathcal{L}(\bm{\alpha}) &=& \prod^{N_\star}_j \int p(D_j|\bm{z}_j) p_\star(\bm{z}_j|\bm{\alpha}) d\bm{z}_j.
\end{eqnarray}
This means that the NPPS function and the stellar parameter distribution can be inferred independently.
The likelihood in Equation (\ref{eq:no_z_NPPS}) has been adopted in previous studies of the occurrence rate 
that do not take into account its dependence on stellar parameters \citep[e.g., ][]{Foreman-Mackey+2014, Fulton+2021}.

\subsection{Ignoring Dependency on Planet Properties}

Let us consider the other extreme case, where we instead ignore the dependency of the NPPS function on the properties of planets. In this case,
\begin{eqnarray}
    {\partial f(\bm{x}|\bm{z}) \over \partial\bm{x}}=0 
    \quad \rightarrow \quad f(\bm{x}|\bm{z}) = {h(\bm{z}) \over \Delta\bm{x}},
\end{eqnarray}
where $h(\bm{z})$ denotes the NPPS as a function of $\bm{z}$ alone, and $\Delta \bm{x}$ is the volume of the domain of $\bm{x}$. The latter domain implicitly defines what is counted as planets in the NPPS $h$.
Then the part of Equation (\ref{eq:DjHj}) relevant for the planets becomes as follows:
\begin{eqnarray}
    &&\left(\prod_{\bm{x}_j\in H_j} \widetilde{f}_j(\bm{x}_j|\bm{z}_j,\bm{\gamma})\right)\exp{\left[-\int \widetilde{f}_j(\bm{x}|\bm{z}_j,\bm{\gamma})d\bm{x} \right]} \nonumber\\
    &=& \left(\prod_{\bm{x}_j\in H_j} \eta_j(\bm{x}_j) \frac{h(\bm{z}_j,\bm{\gamma})}{\Delta\bm{x}}\right) \exp{\left[-\frac{h(\bm{z}_j,\bm{\gamma})}{\Delta\bm{x}} \int \eta_j(\bm{x}) d\bm{x} \right]} \nonumber \\
    &\propto& \left(\prod_{\bm{x}_j\in H_j} \eta_j(\bm{x}_j)\right) \left[h(\bm{z}_j,\bm{\gamma})\right]^{n_j} \exp{\left[-\bar{\eta}_j h(\bm{z}_j,\bm{\gamma})\right]}, \\ \nonumber
\end{eqnarray}
where 
\begin{equation}
    \bar{\eta}_j = {1\over \Delta\bm{x}} \int \eta_j(\bm{x})\,d\bm{x}
\end{equation} 
denotes the detection efficiency $\eta_j(\bm{x})$ averaged over the domain of $\bm{x}$.
The product $\prod_{\bm{x}_j\in H_j} \eta_j(\bm{x}_j)$ is constant when $\eta_j$ is given and so can be ignored in the inference.
Then Equation (\ref{eq:DjHj}) is simplified as:
\begin{eqnarray}
    &&p(D_j,H_j|\bm{\alpha},\bm{\gamma}) \nonumber\\
    &\propto& \int p(D_j|\bm{z}_j)\,p_\star(\bm{z}_j|\bm{\alpha}) \left[h(\bm{z}_j,\bm{\gamma})\right]^{n_j} \exp{\left[-\bar{\eta}_j h(\bm{z}_j,\bm{\gamma})\right]} d\bm{z}_j. \nonumber \\
\end{eqnarray}
This is essentially the likelihood function adopted in \citet{Masuda2022}, who used the binomial distribution (instead of Poisson) to model the fraction of stars showing significant rotational modulation.

\section{Corner plots for the Parameters in the Parametric Inference}

\begin{figure*}[ht!]
    \centering
    \includegraphics[bb=0 0 1128 1151,scale=0.22]{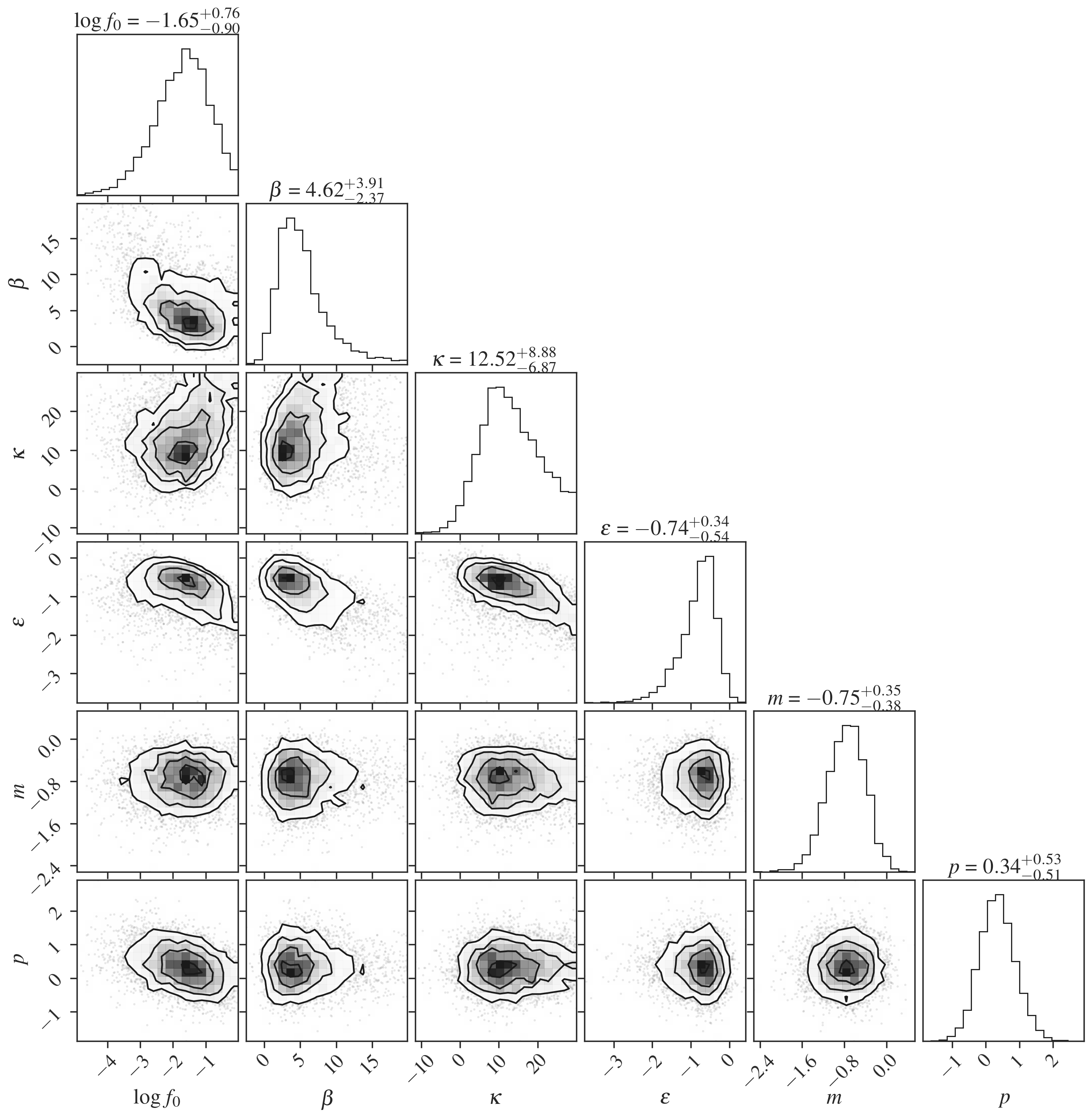}
    \includegraphics[bb=0 0 1128 1151,scale=0.22]{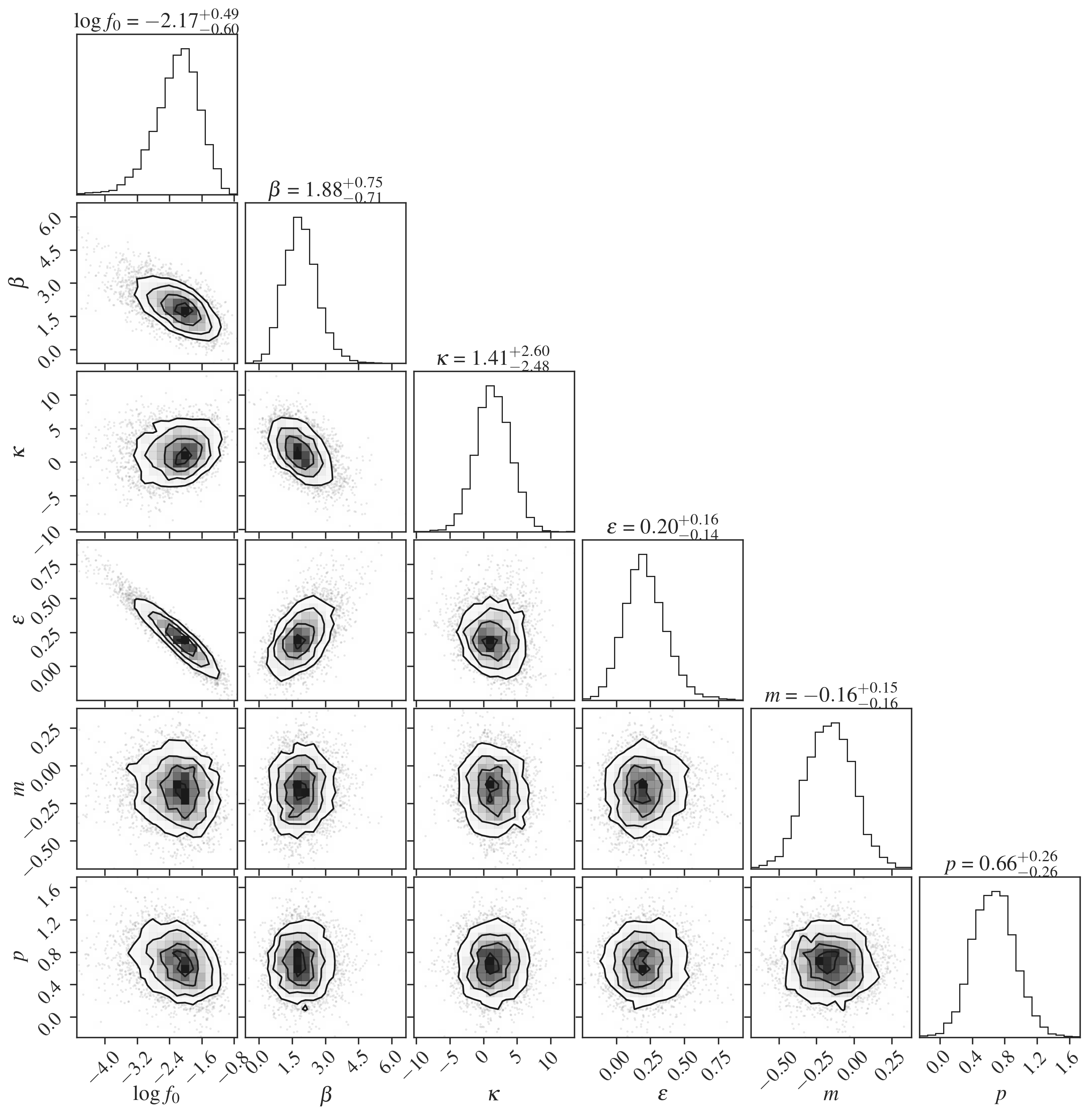}
    \caption{
    Corner plots for the posterior distributions of the parameters in the exponential NPPS model. 
    The left panel is for HJs and the right panel is for CJs. 
    The figure was created using {\tt corner.py} \citep{corner}.
    }
    \label{fig:MCMC_exp}
\end{figure*}

\begin{figure*}[ht!]
    \centering
    \includegraphics[bb=0 0 1128 1151,scale=0.22]{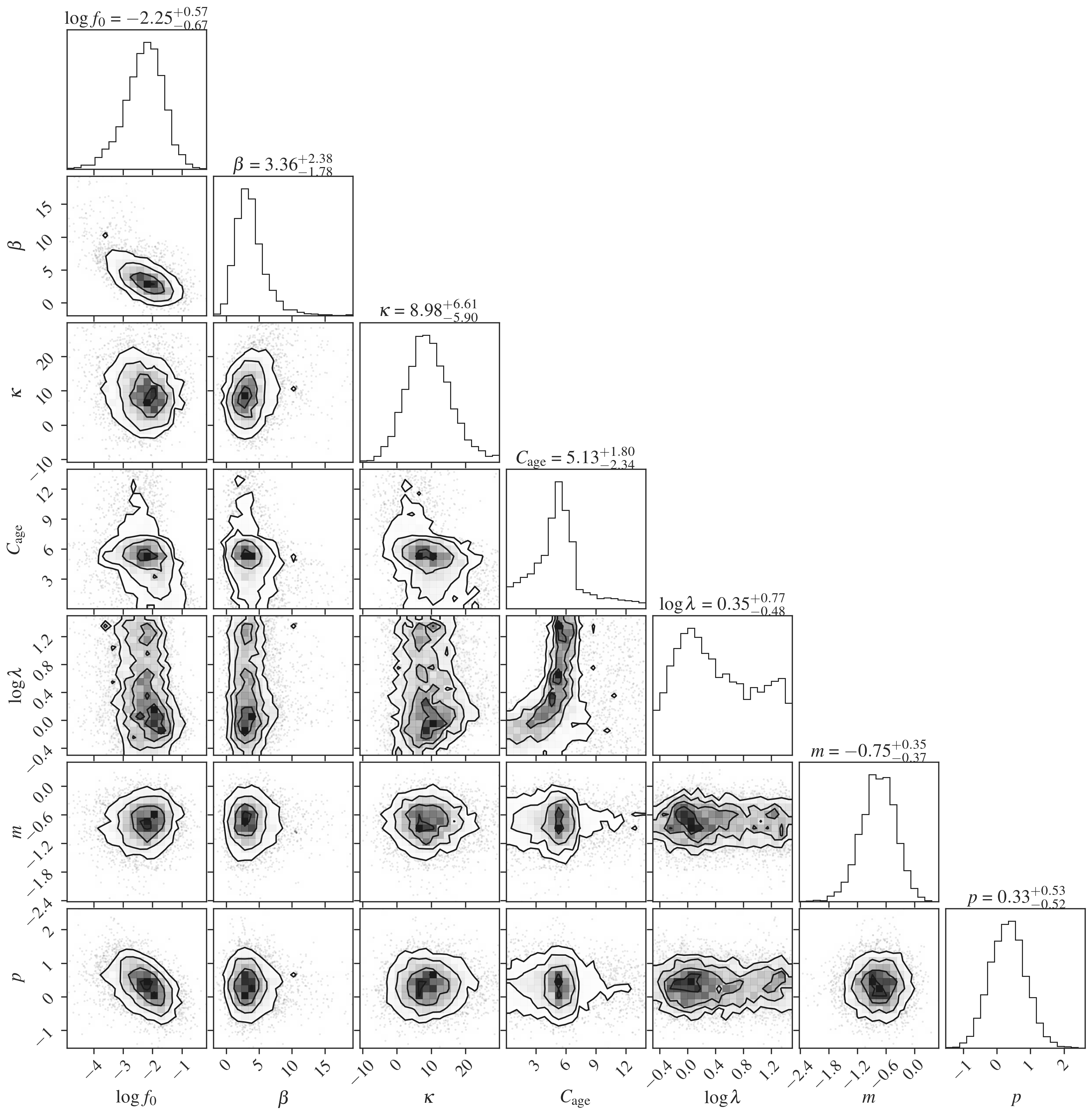}
    \includegraphics[bb=0 0 1128 1151,scale=0.22]{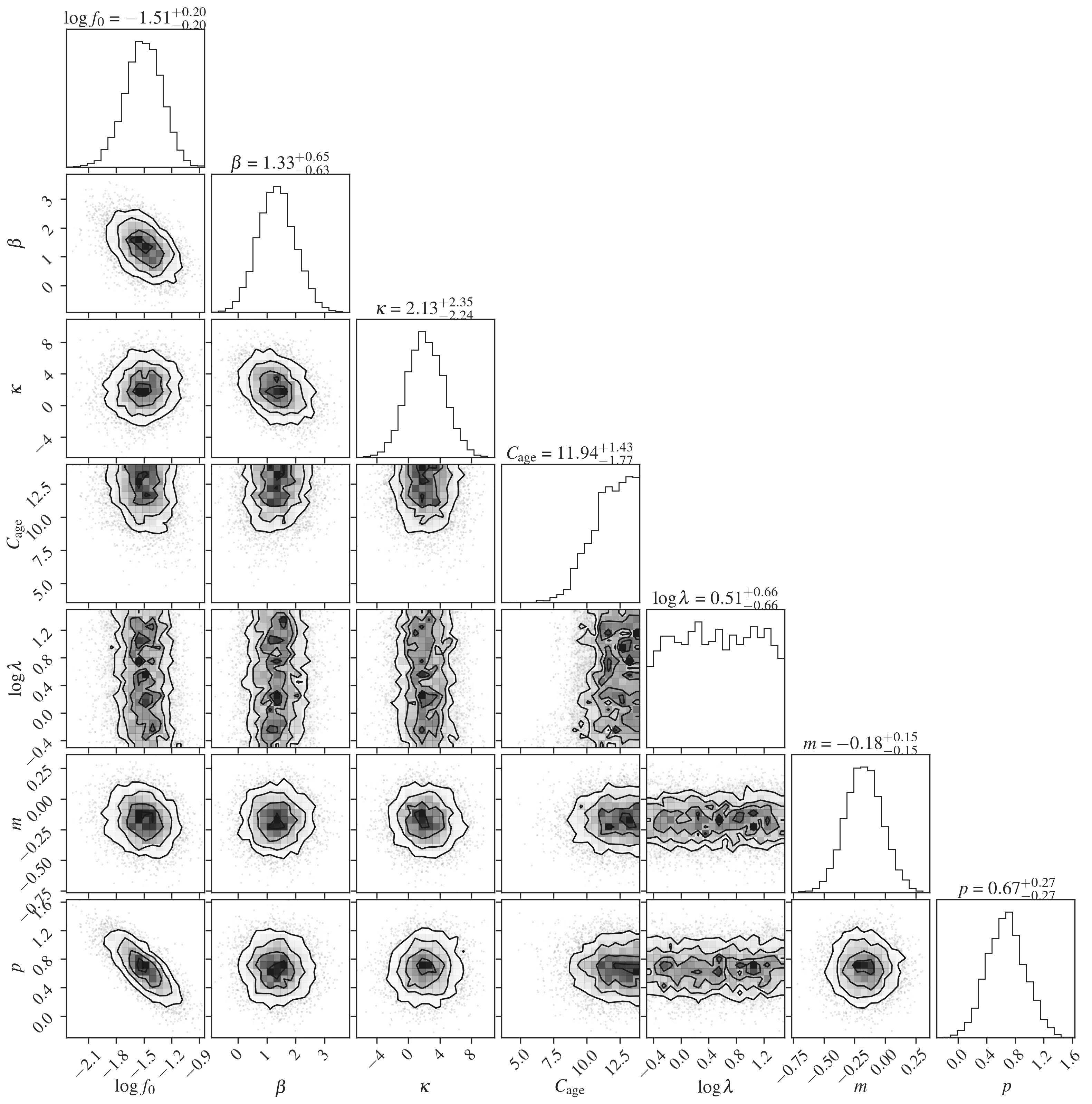}
    \caption{
    Same as Figure \ref{fig:MCMC_exp} but for the sigmoid model.
    }
    \label{fig:MCMC_sig}
\end{figure*}

In Figures \ref{fig:MCMC_exp} and \ref{fig:MCMC_sig}, we present the corner plots for the posterior distributions of the parameters in the NPPS models adopting the exponential and sigmoid forms (Section~\ref{subsec:parametric}), respectively. The left panels show the results for HJs, and the right ones are for CJs.

\bibliographystyle{aasjournal}
\bibliography{0_main}
\end{document}